\documentclass[aps,physrev,superscriptaddress,twocolumn]{revtex4-2}
\usepackage{graphicx}
\usepackage{xcolor}
\usepackage{hyperref}
\usepackage{gensymb}
\usepackage{amsmath}
\hypersetup{colorlinks=true,
            linkcolor=blue,
            citecolor=blue,
            urlcolor=orange}
\usepackage[percent]{overpic} 
\usepackage{xcolor}

\usepackage{xcolor}
\definecolor{darkgreen}{rgb}{0.0, 0.5, 0.0}

\usepackage{ulem}

\begin{document}


\title{
Non-Fermi-liquid behavior and Fermi-surface expansion \\induced by van Hove-driven ferromagnetic fluctuations:\\ {the} D-TRILEX analysis} 

\author{Ilia S. Dedov}
\affiliation{Center for Photonics and 2D Materials, Moscow Institute of Physics and Technology, Institutsky lane 9, Dolgoprudny, 141700, Moscow region, Russia}

\author{Andrey A. Katanin}
\affiliation{Center for Photonics and 2D Materials, Moscow Institute of Physics and Technology, Institutsky lane 9, Dolgoprudny, 141700, Moscow region, Russia}
\affiliation{M. N. Mikheev Institute of Metal Physics of the Ural Branch of the Russian Academy of Sciences, S. Kovalevskaya Street 18, 620990 Yekaterinburg, Russia}

\author{Evgeny A. Stepanov}
\affiliation{CPHT, CNRS, \'Ecole polytechnique, Institut Polytechnique de Paris, 91120 Palaiseau, France}
\affiliation{Coll\`ege de France, 11 place Marcelin Berthelot, 75005 Paris, France}


\begin{abstract}
We consider the electronic and magnetic properties of the Hubbard model on a square lattice with the Fermi level near van Hove singularity and the ratio of the next-nearest-neighbor and nearest-neighbor hoppings ${t'/t=-0.45}$, which favors the ferromagnetic instability. 
We find, that a self-consistent consideration of the ferromagnetic fluctuations within the \mbox{D-TRILEX} approach results in the splitting of the electronic spectral function at low temperatures.
This splitting exhibits only a weak momentum dependence, and only one of the split bands crosses the Fermi level. 
As a result, the Fermi surface itself remains unsplit, but its area increases, reflecting the presence of non-Fermi-liquid electronic excitations.
We show that both the self-consistent account of the non-local contributions to the electronic self-energy and the proper {treatment} of electron interaction vertices in \mbox{D-TRILEX} are important to obtain this behavior.
\end{abstract}

\maketitle


\section{\label{sec:level1}Introduction}

The non-Fermi liquid behavior due to magnetic fluctuations was initially discussed in the context of cuprates and other high-temperature superconductors starting from the concept of nearly antiferromagnetic Fermi liquid (see, e.g., Refs.~\cite{Chubukov, Pines1, Pines, Tremblay, SPS}). In particular, it was emphasized that at finite temperature in two dimensions in the vicinity of an antiferromagnetic instability the real and imaginary parts of the electronic self-energy on the real frequency axis behave as  $\Re\Sigma(\nu)\propto (T\ln\xi)/(\nu-\epsilon_{\mathbf{k+Q}})$, $\Im\Sigma(0)\propto -T\xi$ (where $\nu$ is the real energy, $\epsilon_{\mathbf k}$ is the dispersion, $\xi$ is the correlation length, and ${\mathbf Q}$ is the antiferromagnetic wave vector). These dependencies can drive non-Fermi-liquid behavior, and in the proximity of an ordered state, can yield (weak-coupling) pseudogap in the electronic spectrum. Later, these scenarios of the pseudogap were complemented and extended within the strong-coupling analysis; see, e.g., Refs.~\cite{PRL2024, PG0, PG1, PG2, PG3}.

While at first glance the situation regarding ferromagnetic instability appears different, a weak-coupling analysis for this instability, as shown in Refs.~\cite{KK,KTr,QS4}, also yields the above-cited result for the self-energy with ${{\bf Q}=0}$. However, this leads to a quasi-splitting of the Fermi surface above the ferromagnetically ordered state (see also Ref.~\cite{QS3}). Similarly to the case of antiferromagnetic fluctuations, this analysis (including also the results of the functional renormalization group (fRG) approach~\cite{KK}) requires studying the effect of {\it local} spin correlations, which arise at moderately strong Coulomb interactions and are essential for describing phenomena occurring in realistic systems in the vicinity of ferromagnetic instability.

A prototypical example of systems 
with potential importance of both, local and non-local magnetic correlations {are perovskite materials with partially filled $t_{2g}$ orbitals~\cite{SrRu0, SrRu1, HundMetal, SrRu2}, such as} Sr$_2$RuO$_4$.
The proximity of the Fermi level to a van Hove singularity (vHS) in these materials leads to an enhanced density of states, promoting ferromagnetic fluctuations (which coexist with spin-density-wave fluctuations arising from the nesting of other Fermi surface sheets),
and resulting in non-Fermi-liquid behavior at low temperatures~\cite{SrRuFM, Steffens}.
These fluctuations and non-Fermi-liquid behavior can be further enhanced by shifting the Fermi level closer to the vHS via substitution of ruthenium by lanthanum, corresponding to  electron doping~\cite{Kikugawa}, yet the system remains paramagnetic. Also, strong local correlations and quite a small Kondo temperature $T_K\sim 25 K$ was discussed for Sr$_2$RuO$_4$~\cite{Mravlje, SrRuFM, Kugler}. 
This resilience against ordering, while hosting strong fluctuations, highlights the complex interplay between electronic correlations, Fermi surface topology and non-Fermi-liquid behavior \cite{SrRu2}. Materials like Sr$_2$RuO$_4$
provide a unique platform to isolate these effects and study a possibility of electronic spectrum splitting in paramagnetic phase. 
We note that recently such a splitting was  observed in the layered van der Waals material CrTe$_2$ \cite{CrTe2}.
The splitting of electronic spectrum by ferromagnetic fluctuations was also discussed within non-local extensions of dynamical mean field theory (DMFT) in Na$_x$CoO$_2$ \cite{Wilhelm} and recently in CrTe$_2$ and some three-dimensional magnets \cite{Katanin}. 


While these works have provided valuable insights into fluctuation–driven spectral anomalies, the role of the non-local self-energy effects near a vHS in the presence of ferromagnetic correlations 
remains 
not fully understood. 
In this work, we apply the dual triply irreducible local expansion (\mbox{D-TRILEX}) method~\cite{DTRILEX1, DTRILEX2, DTRILEX3}, which provides a consistent route to incorporate non-local correlations on top of DMFT via the three-leg vertex functions, to a single-band model that features both a van Hove singularity in the electronic spectrum and ferromagnetic fluctuations.
This framework enables us to investigate how ferromagnetic fluctuations near the van Hove singularity influence the self-energy and 
momentum-resolved spectral functions.
By contrasting \mbox{D-TRILEX} to the purely local DMFT solution, we clarify the crucial role of momentum-dependent fluctuations in driving non-Fermi-liquid behavior in the two-dimensional Hubbard model. 
To further elucidate the impact of self-consistency, we also compare the obtained spectral functions to those in the non-self-consistent dynamical vertex approximation (D$\Gamma$A) \cite{DGA,DGA1}, which was also recently applied to the problem under consideration in Ref. \cite{QS4}. 
This comparison allows us to assess how the two methods reproduce the underlying band structure and spectral weight distribution, demonstrating that \mbox{D-TRILEX}  {provides a more physically consistent description of the bands}. We also find that non-local corrections strongly enhance quasiparticle damping and give rise to pronounced non-Fermi-liquid features even in the absence of symmetry breaking.

The paper is organized as following. In Section~II, we summarize the theoretical approach utilized in this work. In Section~III, we present the obtained phase diagram (Sect.~III\,A), self-energies (Sect.~III\,B.1) and spectral functions (Sect.~III\,B.2), showing non-Fermi-liquid behavior of electronic degrees of freedom. In Section~IV, we present our conclusions and outlook.


\vspace{-0.2cm}
\section{\label{sec:methods}The model and method}

For the numerical analysis we study the Hubbard model on a square-lattice:
\begin{equation}
H = -\sum_{ij\sigma} t_{ij} c^\dagger_{i\sigma} c_{j\sigma} + U \sum_i n_{i\uparrow} n_{i\downarrow}
\end{equation}
with the nearest-neighbor hopping $t$, the next-nearest-neighbor hopping ${t'=-0.45t}$, and the Coulomb interaction ${U=4t}$; in the following we put ${t=1}$. We investigate the system away from half-filling, in the density range ${0.4 < n < 0.6}$, which corresponds to the Fermi level being close to the van Hove singularity (vHS). 
At low temperatures strong ferromagnetic fluctuations emerge due to the vHS, according to the Stoner criterion. 
This model allows one to capture the interplay of non-local electronic correlations and van Hove physics.

To {investigate the} self-energy of electrons in the vicinity of the vHS, we first employ DMFT. While DMFT provides a reliable description of local dynamical correlations, it neglects non-local corrections. This limitation is particularly severe near a vHS, where enhanced scattering processes generate strong momentum-dependent self-energy effects. 

To go beyond the purely local DMFT framework, we use the dual triply irreducible local expansion (\mbox{D-TRILEX}) method~\cite{DTRILEX1, DTRILEX2, DTRILEX3}. This approach combines the strengths of DMFT in describing local dynamical correlations with the ability of \mbox{D-TRILEX} diagrammatic to efficiently account for the non-local contributions to the self-energy and susceptibilities induced by the van Hove singularity. Specifically,
\mbox{D-TRILEX} 
incorporates non-local correlations by self-consistently computing the self-energy $\tilde{\Sigma}(i\nu_n, \mathbf{k})$ and polarization operator $\tilde\Pi(i\omega_n,\mathbf{k})$ in an effective ``dual'' space via the three-point (triangular) vertex functions $\Lambda_{\omega \nu}$ obtained from the DMFT impurity problem:
\begin{align}
\tilde{\Sigma}_{\mathbf{k}\nu} &=
- \sum_{\mathbf{q},\omega,\varsigma} 
\Lambda^{\varsigma}_{\nu\omega} \, \tilde{G}_{\mathbf{q}+\mathbf{k},\nu+\omega,\sigma} \, \tilde W_{\mathbf{q}, \omega}^{\varsigma} \, \Lambda^{\varsigma}_{\nu+\omega,-\omega}\,, \\
\tilde{\Pi}_{ \mathbf{q} \omega}^\varsigma &=
2 \sum_{\mathbf{k}}
\Lambda_{\nu+\omega,-\omega}^\varsigma \, 
\tilde{G}_{\mathbf{k}\nu} \, 
\tilde{G}_{\mathbf{k}+\mathbf{q}, \nu+\omega} \,
\Lambda_{\nu\omega}^\varsigma\,,
\end{align} 
where ${\varsigma=c,s}$ denotes the charge and spin channels, 
$\tilde{G}_{{\mathbf k}\nu}=(\tilde{\mathcal{G}}_{{\mathbf k}\nu}^{-1}-\tilde{\Sigma}_{\mathbf{k}\nu})^{-1}$ is the dressed dual Green's function, and $\tilde W_{\mathbf{q} \omega}^{\varsigma}=(\tilde{\mathcal{W}}_{\omega}^{\varsigma\,-1}-\tilde{\Pi}^\varsigma_{\mathbf{q}\omega})^{-1}$ is the renormalized dual interaction, which can be obtained from the Dyson equations, using bare propagators:
\begin{align}
\tilde {\mathcal{G}}_{\mathbf{k}\nu} &= [(\epsilon_\mathbf{k} - \Delta_\nu)^{-1} - {g}_{\nu}]^{-1}\,,\label{Eq1}\\
\tilde{\mathcal{W}}_{\omega}^\varsigma  &= \mathcal{W }_{\omega}^\varsigma  - \frac{1}{2}U^\varsigma \,.
\label{Eq2}    
\end{align}
In Eq.~\eqref{Eq1}, $\Delta_\nu$ and {$g_{\nu}$} are respectively the hybridization function and the Green's function of the DMFT impurity problem. In Eq.~\eqref{Eq2}, $\mathcal{W}^{\varsigma}_{\omega} 
=  (U^{\varsigma\,-1} - \Pi^{\varsigma}_{\omega})^{-1}$ is the EDMFT-like renormalized interaction dressed in the local polarization operator of the DMFT impurity problem $\Pi^{\varsigma}_{\omega}$, and we use $U^{c/s}=\pm U$.
{The physical electronic Green's function $G({\bf k},i\nu)$, self-energy $\Sigma({\bf k},i\nu)$, and spin susceptibility $X^{s}({\bf q},i\omega)$ investigated in this work are further obtained via the exact relations between the dual and lattice quantities. The detailed description of the method can be found in Ref.~\cite{DTRILEX3}.}

The \mbox{D-TRILEX} approach is particularly suitable for studying the effect of van Hove singularity, where non-local fluctuations strongly modify the self-energy. 
An important advantage of \mbox{D-TRILEX} over other diagrammatic extensions of DMFT (such as D$\Gamma$A ~\cite{DGA, DGA1, SCDGA} or dual fermion/boson approaches~\cite{PhysRevB.77.033101, BRENER2020168310, Rubtsov20121320, PhysRevB.90.235135, PhysRevB.93.045107, PhysRevB.100.165128})  lies in its computational efficiency. 
Specifically, the method employs a partially bosonized approximation ~\cite{DTRILEX1}, greatly reducing the computational cost of the diagrammatic expansion while retaining the essential physics of long-range fluctuations~\cite{PhysRevLett.127.207205, stepanov2021coexisting, PhysRevLett.129.096404, PhysRevResearch.5.L022016, PRL2024, PhysRevLett.132.226501, PhysRevB.110.L161106, stepanov2023charge, j6bj-gz7j, SrRu2, arXiv.2502.08635}.
This allows us to perform self-consistent calculations on fine momentum grids and to explore parameter regimes where the non-local correlations play a decisive role.

For DMFT calculations of local self-energy and triangular vertices we use segment continuous time Monte-Carlo (CT-QMC)~\cite{CT-QMC1} hybridization expansion~\cite{CT-QMC2} \mbox{iQIST} solver~\cite{iQIST}, which allows us to considerably reduce the stochastic noise and obtain the details of the frequency dependence of the vertex functions over a sufficiently broad frequency range. The momentum-resolved spectral function  $A(\mathbf{k},\omega) = -\frac{1}{\pi}\,\mathrm{Im}\,G(\mathbf{k},\omega)$ in DMFT and \mbox{D-TRILEX} is obtained from the Matsubara Green’s function $G(\mathbf{k},i\nu)$ via analytic continuation using the maximum entropy method, as implemented in the \texttt{ana\_cont} package~\cite{ana_cont}. From the self-energies we also calculate quasiparticle damping $\gamma_k $.

\section{Numerical results}

\subsection{Phase Diagram}

\begin{figure}[b!]
    \centering
    \includegraphics[width=0.9\linewidth]{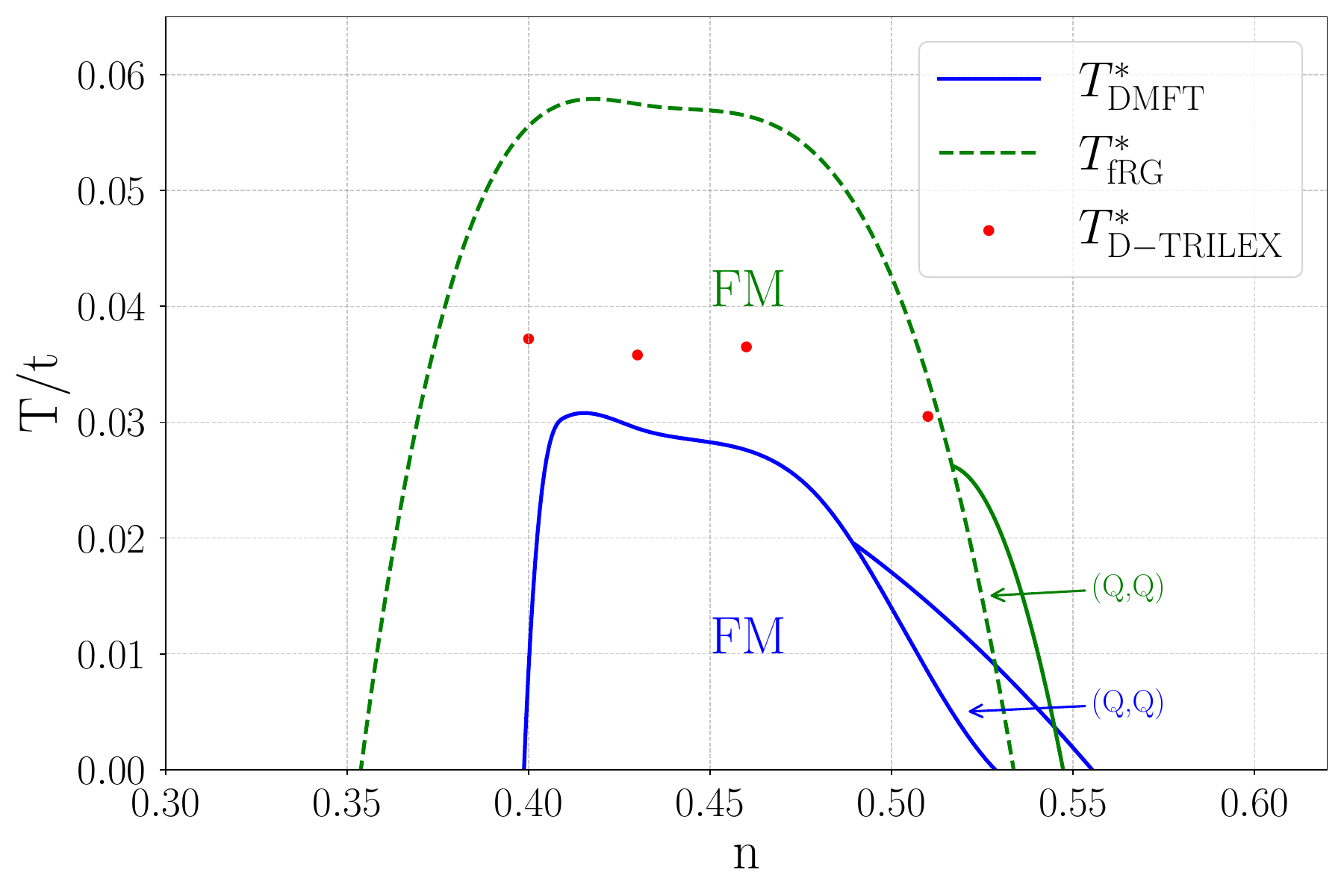}
    \caption{
Phase diagram in the $n$–$T$ plane obtained within fRG (green line), DMFT (blue line), and \mbox{D-TRILEX} (red points). The $(Q,Q)$ region indicates an incommensurate magnetic phase, while FM denotes the ferromagnetic one.}

    \label{fig:diag}
\end{figure}

In order to calculate the transition temperatures to the ferromagnetic state, illustrated in Fig.~\ref{fig:diag}, we calculate the static (${\omega=0}$) \mbox{D-TRILEX} spin susceptibility {$X^{s}_{\bf q} = \langle S^{z}_{\bf q} \,S^{z}_{\bf -q}\rangle$}
and extrapolate the inverse of it to zero. 
The obtained temperatures are compared with the fRG transition temperatures~\cite{fRG2011}, which do not account for quasiparticle damping, and 
to those in DMFT approach. The spin susceptibilities in DMFT are obtained from the respective Bethe-Salpeter equation {written in terms of the DMFT Green's functions and the numerically exact local four-point vertices of the impurity problem} (we also use the corrections for finite frequency box in Bethe-Salpeter equations~\cite{My_BS}). Since all considered approaches violate Mermin-Wagner theorem, the obtained magnetic transition temperatures should be considered as the crossover temperatures $T^*$ to the renormalized classical regime with strong magnetic correlations, characterized by exponentially large correlation length.

In DMFT, we obtain a $T^*$ that is roughly half of that predicted by fRG; the suppression of the crossover temperature $T^*$ is induced by local self-energy effect. The region of the states which have a tendency to magnetic ordering also shrinks in DMFT on the side of small concentrations ${n\sim 0.35}$. 
Both DMFT and fRG consistently predict a strong tendency toward incommensurate magnetism at higher $n$. 
In \mbox{D-TRILEX}, we find somewhat higher crossover temperatures than those obtained within DMFT, 
which we attribute to the partially bosonized treatment of the four-point vertex function, approximated by the three-point vertices connected via bosonic fluctuations.
Yet, similarly to DMFT, the obtained crossover temperatures are much lower than those obtained in fRG approach, due to the inclusion of self-energy effects. 
Moreover, the self-consistent treatment of the non-local self-energy in \mbox{D-TRILEX} leads to a significantly weaker tendency toward incommensurate order compared to DMFT.

\subsection{\label{sec:level1}Analysis of non-Fermi-liquid behavior and Fermi surface expansion}
\subsubsection{\label{sec:citeref}Self-energies}
\begin{figure}[b]
    \centering
    \includegraphics[width=0.49\linewidth]{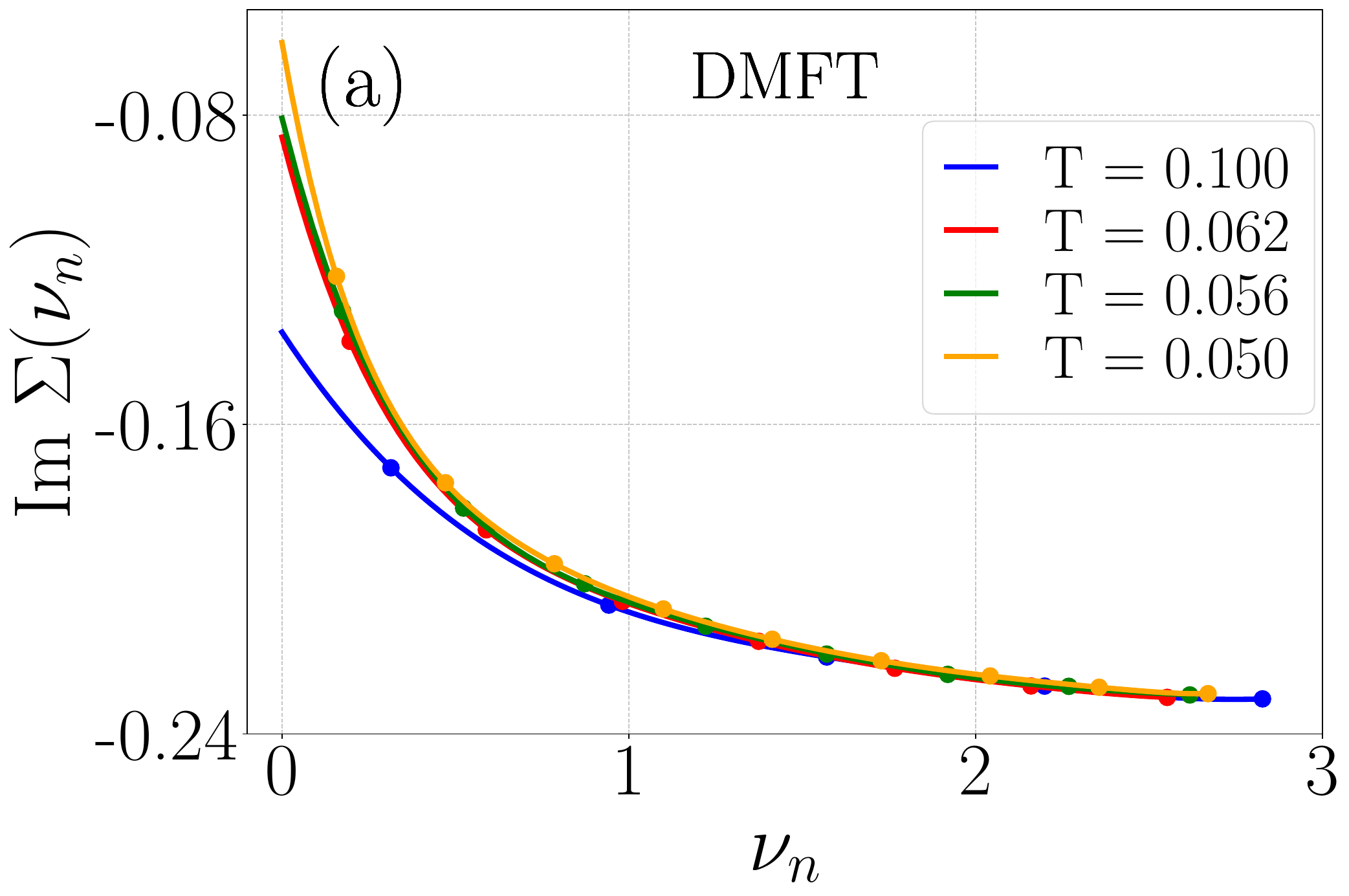}
    \includegraphics[width=0.49\linewidth]{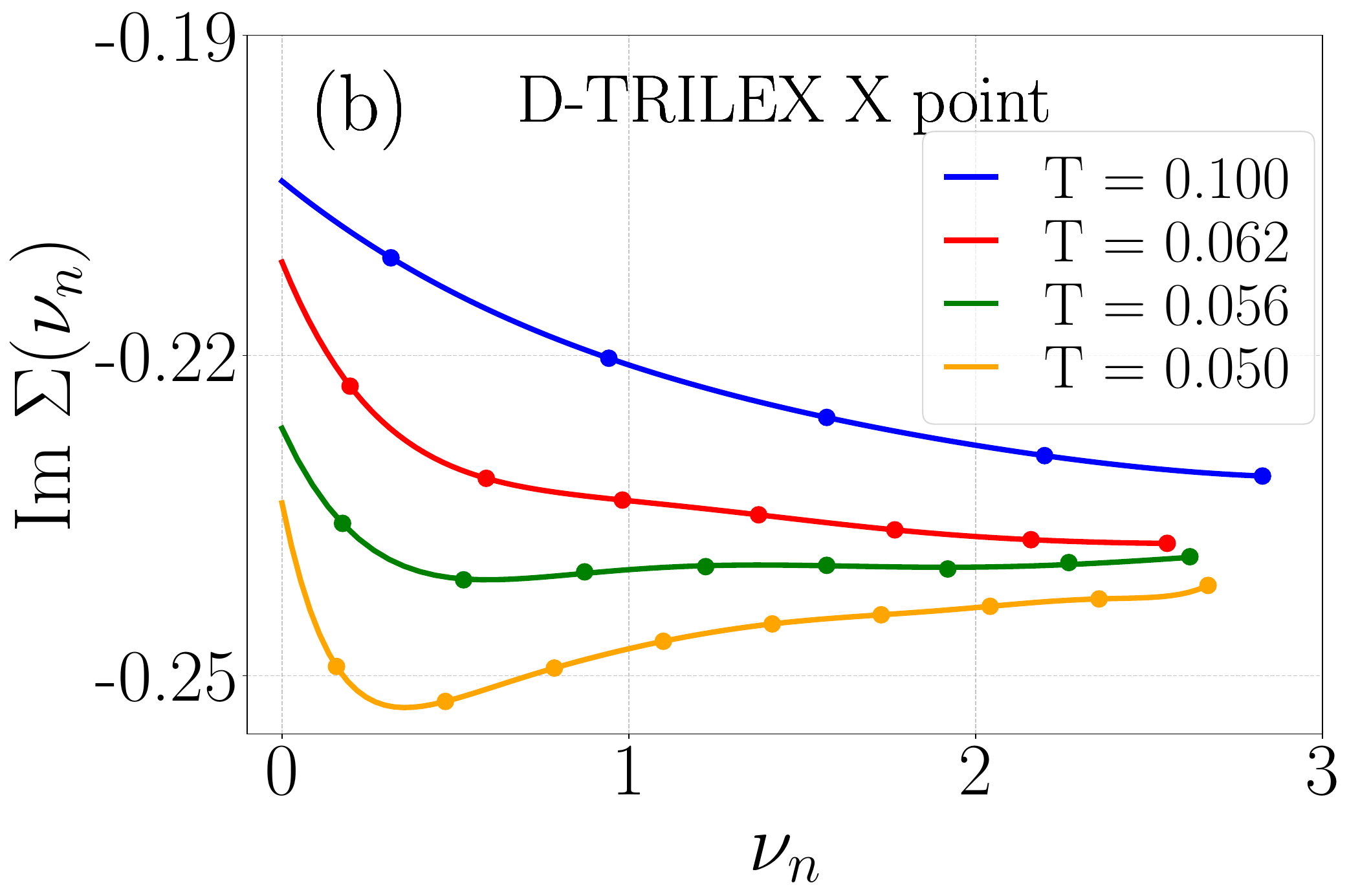}
     \includegraphics[width=0.49\linewidth]{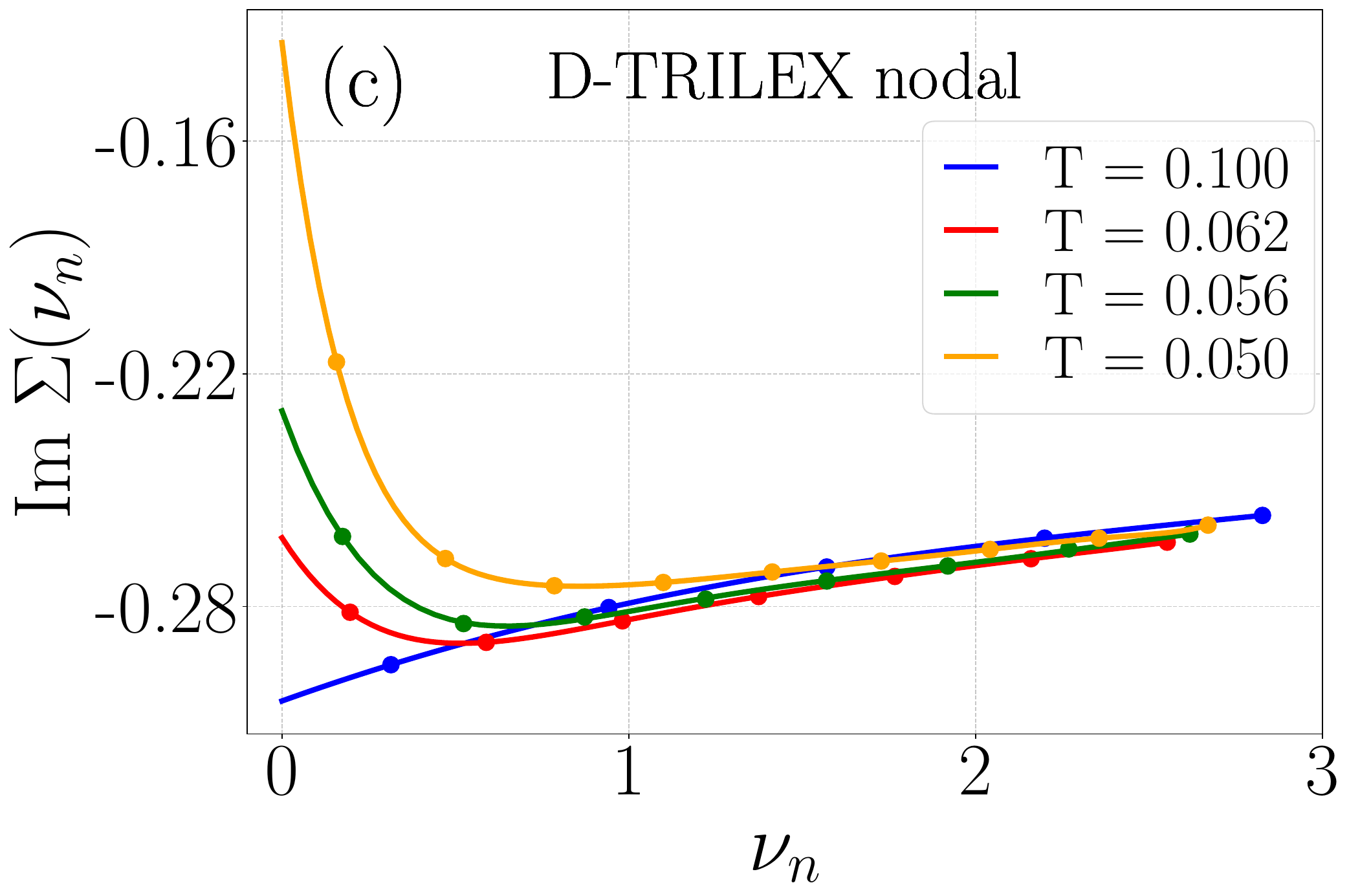}
    \includegraphics[width=0.49\linewidth]{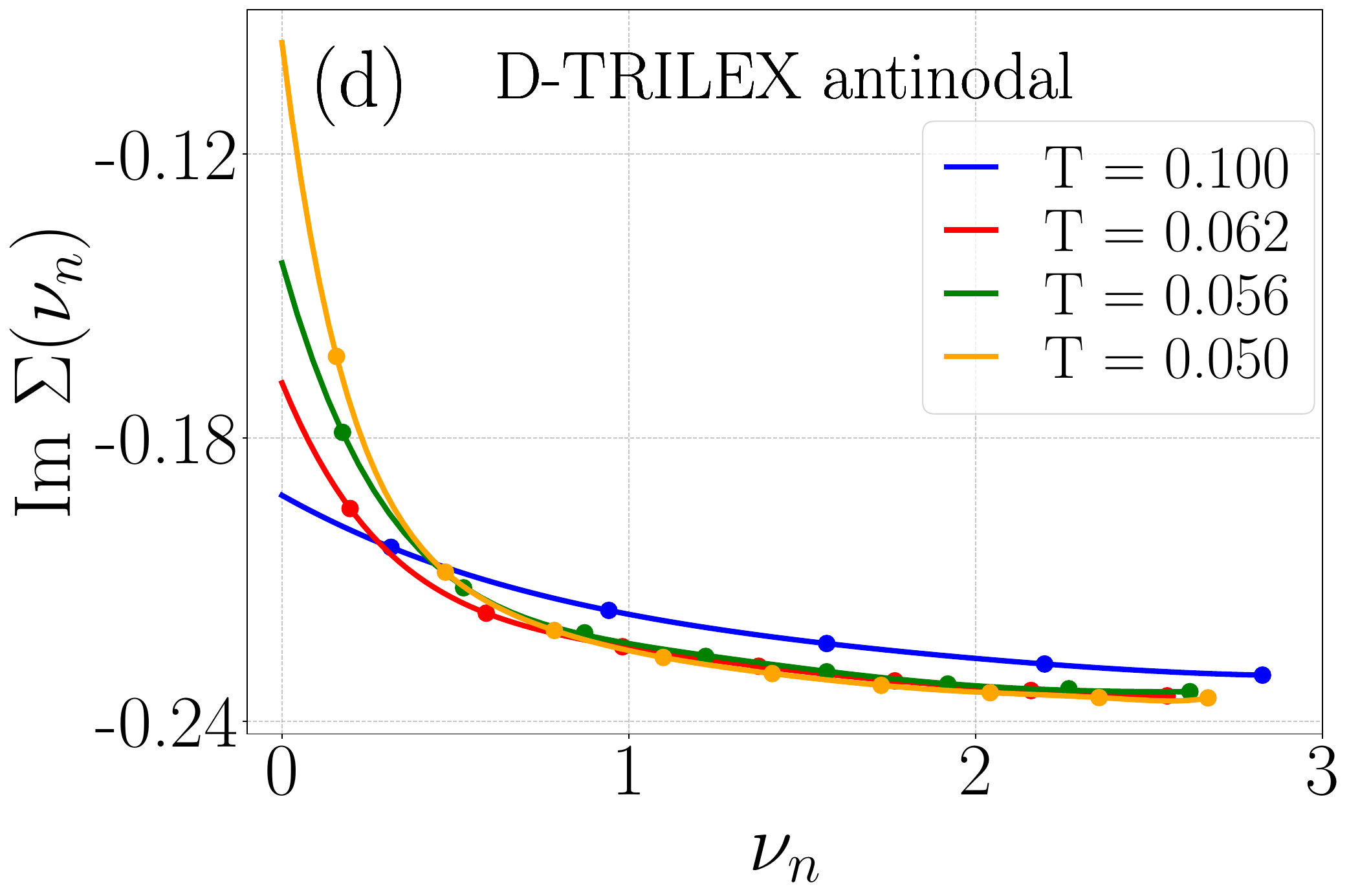}
    \includegraphics[width=0.49\linewidth]{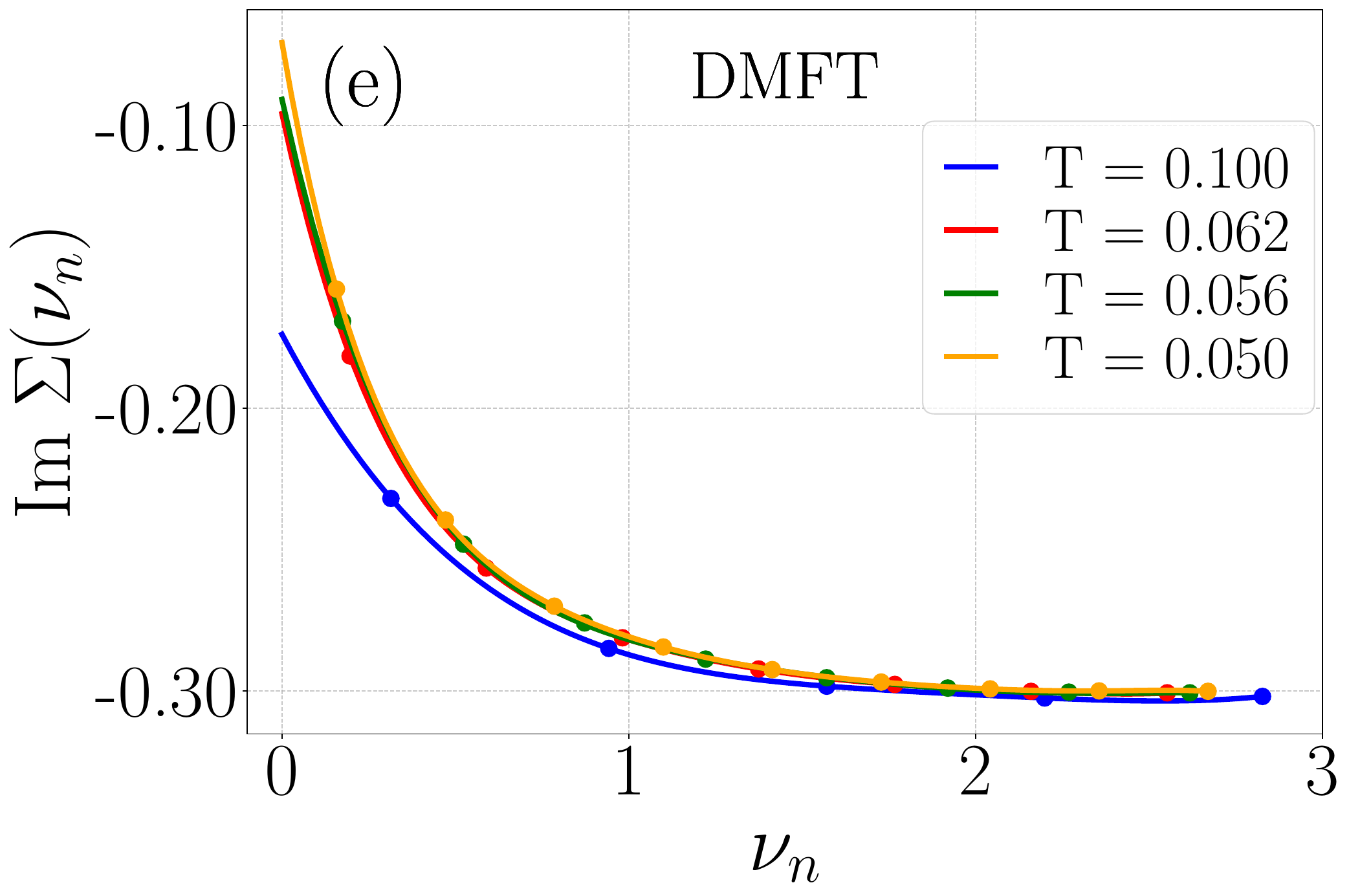}
    \includegraphics[width=0.49\linewidth]{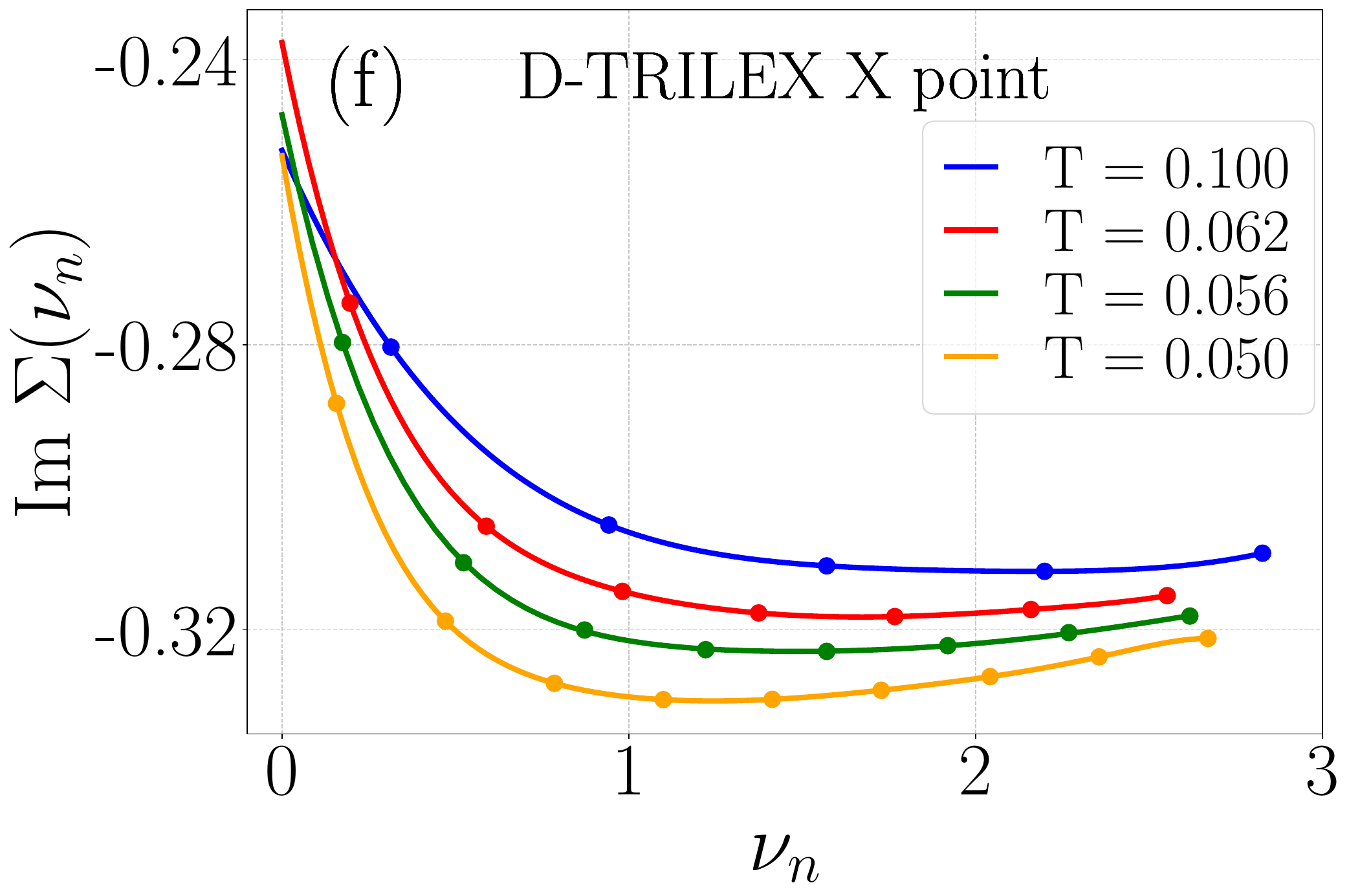}

    \includegraphics[width=0.49\linewidth]{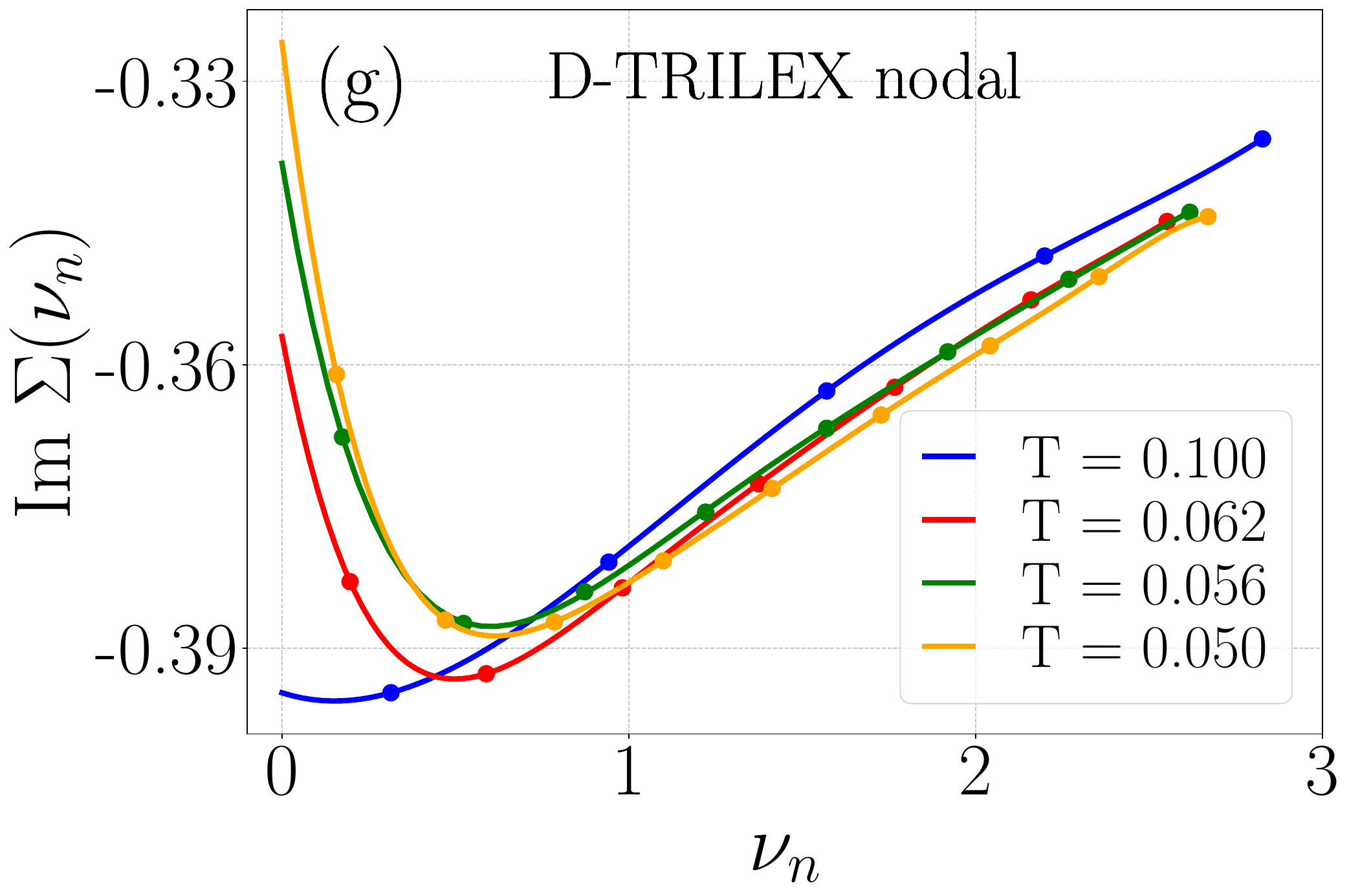}
    \includegraphics[width=0.49\linewidth]{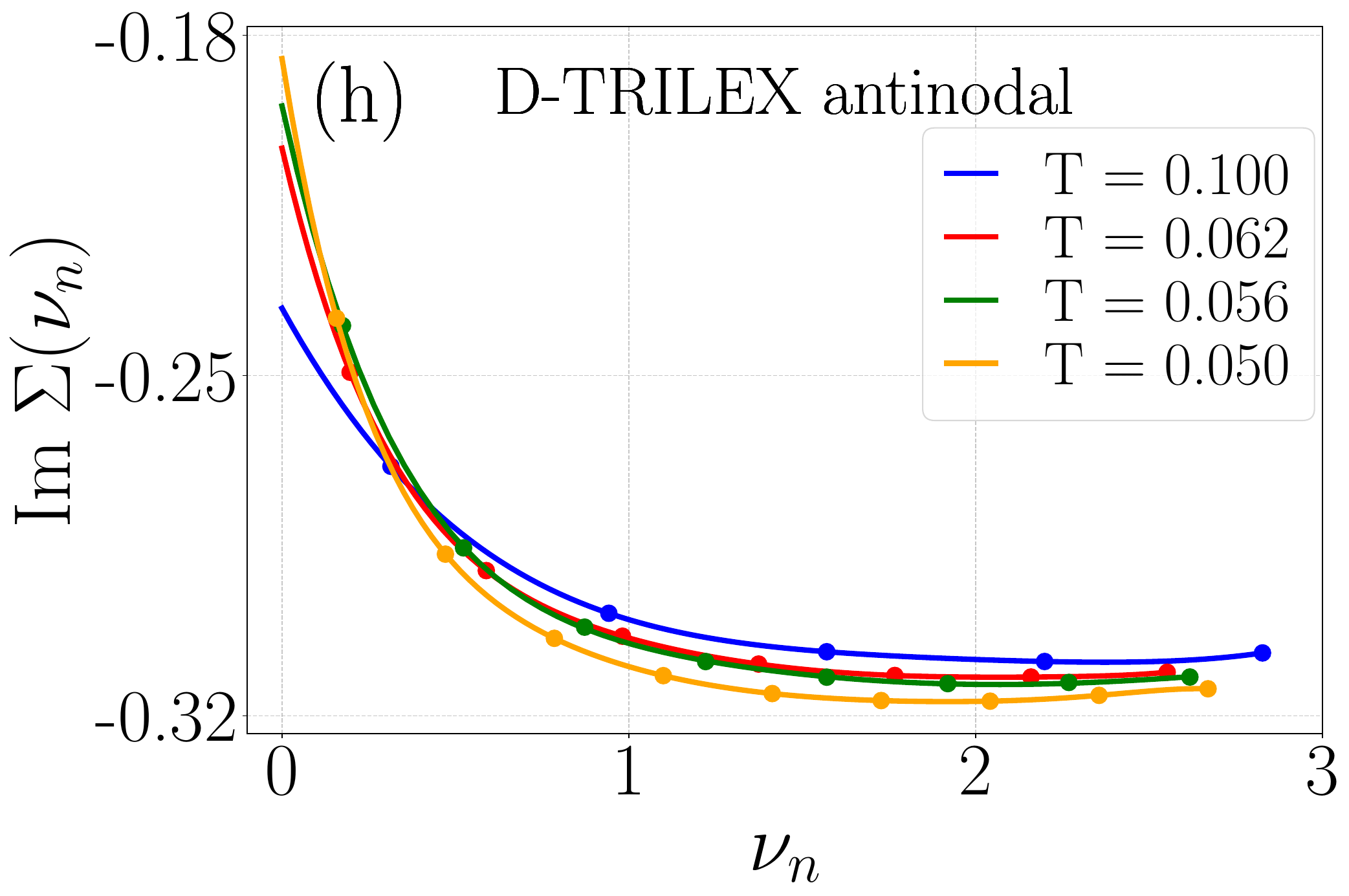}

    \caption{ Self-Energies on Matsubara frequencies at $n = 0.43$ (a-d) and $n=0.52$ (e-h). (a,e) DMFT, (b,f)  \mbox{D-TRILEX} X-point, (c,g) \mbox{D-TRILEX} nodal point, (d,h) \mbox{D-TRILEX} anti-nodal point.  The data have been extrapolated to zero frequency by a polynomial fit.}
    \label{fig:Sigma_n_0.43}
\end{figure}

%

\begin{figure}[t]
    \centering
    \includegraphics[width=0.95\linewidth]{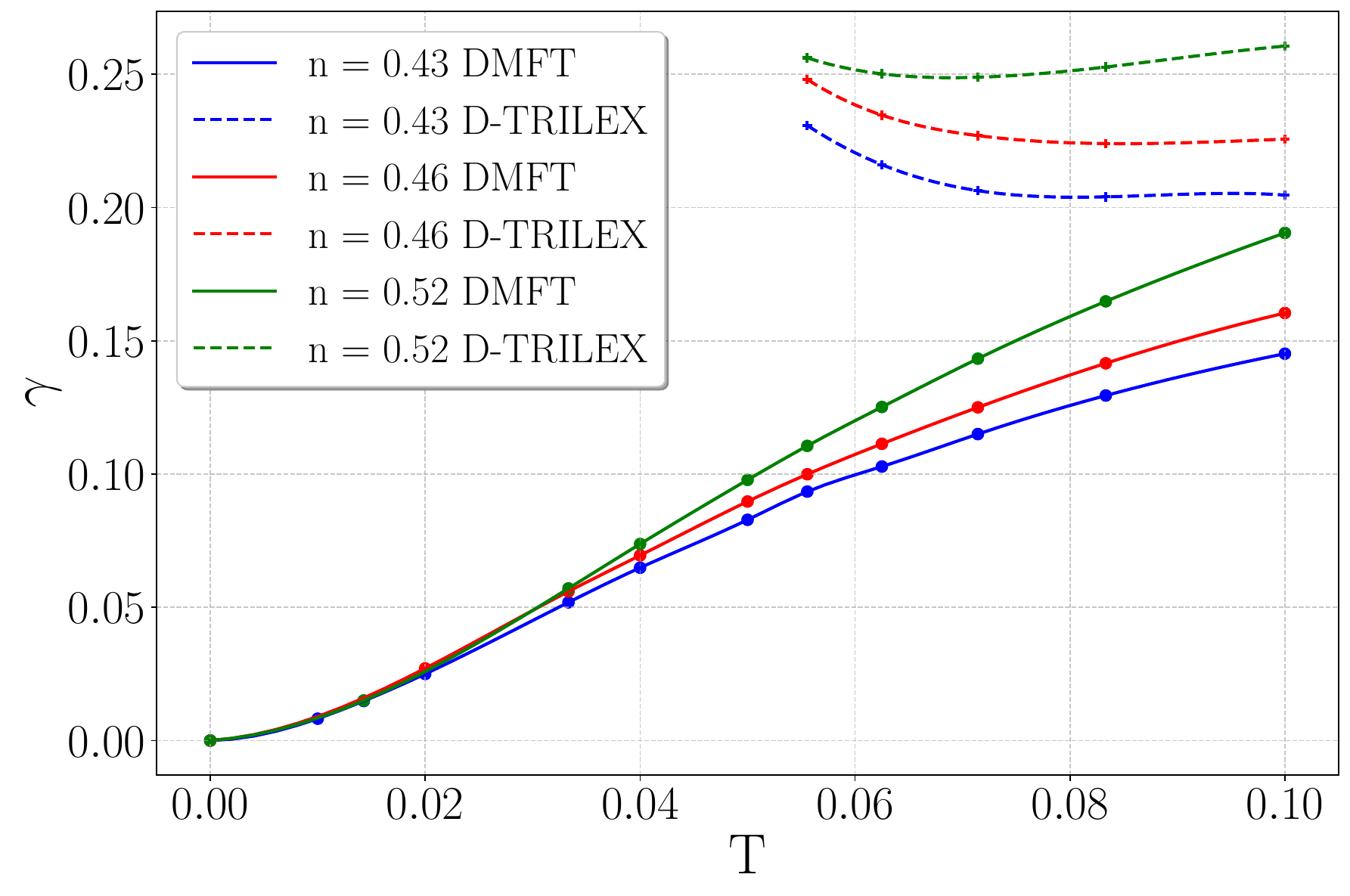}
    \caption{Quasiparticle damping $\gamma =  -\mathrm{Im}\Sigma_{{\mathbf k},\nu = 0}$ in \mbox{D-TRILEX}  at ${\mathbf k}=X=(\pi,0)$ (dashed lines) and DMFT (solid lines) as functions of temperature for $n = 0.43$, $0.46$ and $0.53$}
    \label{fig:gamma}
\end{figure}

We next consider the local self-energies obtained within DMFT and \mbox{D-TRILEX} for ${{0.05}t < T < 0.1t}$, i.e., above the transition temperature, and for fillings ${n = 0.43}$ and ${n=0.52}$ close to the van Hove singularity. The corresponding result is shown in Fig.~\ref{fig:Sigma_n_0.43}. 
In DMFT we find that the quasiparticle damping $\gamma_{\mathbf k}= -\mathrm{Im}\Sigma_{{\mathbf k},\nu = 0}$, {shown in Fig.~\ref{fig:gamma}}, decreases with lowering the temperature and eventually shows a recovery of the expected Fermi-liquid-like $T^2$ scaling. 
For higher temperatures we find a deviation from this scaling law. 
Importantly, the $T^2$ behavior of the self-energy is obtained only at the temperatures much lower than the DMFT transition temperature ${T < T^* \approx 0.03}$. Moreover, the absolute values of $\gamma$ increase with density as the system approaches half-filling. 

In \mbox{D-TRILEX} the self-energies at the van Hove singularity point ${\text{X}=(\pi,0)}$ (Fig.~\ref{fig:Sigma_n_0.43}\,b) exhibit pronounced non-Fermi-liquid behavior, with substantially larger quasiparticle damping compared to DMFT, which further increases as the temperature is lowered.
Furthermore, the \mbox{D-TRILEX} results show a strong momentum differentiation. 
The momenta corresponding to the Fermi surface (determined by largest spectral weight, see below), along the diagonal (``nodal'' direction) and closest to the X point (``antinodal'' direction), still display a sufficiently large quasiparticle damping (Figs.~\ref{fig:Sigma_n_0.43}\,c,\,d,\,g,\,h). 
Notably, $\gamma$ exhibits a temperature trend opposite to that at the X point, slightly decreasing with lowering temperature at both the nodal and antinodal points. Among these two points, the quasiparticle damping is largest at the nodal point and smallest at the antinodal one.

At the X point, the temperature dependence of $\gamma$ in \mbox{D-TRILEX} is extremely weak at high temperatures (see Fig. \ref{fig:gamma}), showing only a slight increase upon lowering $T$, opposite to the DMFT trend. In addition, the absolute values of quasiparticle damping are almost twice as large as in DMFT. These findings demonstrate that non-local corrections to the self-energy associated with the van Hove singularity drastically enhance non-Fermi-liquid behavior and prevent the system from approaching the Fermi-liquid regime as the temperature decreases.

For additional insight into the self-energies, we analyze Im$\Sigma(\nu) - \alpha\nu$ as a function of $\nu^2 - (\pi T)^2$ (see Appendix).  According to Fermi-liquid theory,  Im$\Sigma(\nu) - \alpha\nu \propto \nu^2 - (\pi T)^2$ \cite{FirstFreq},  where the term $\alpha\nu$ originates from real part of self-energy on real axis and is therefore subtracted.  
In DMFT this criterion is fulfilled only at low temperatures  $T < 0.02$.
As the non-local corrections, incorporated in \mbox{D-TRILEX}, enhance non-Fermi-liquid behavior at the X point, this relation is completely violated. At the same time, both nodal and antinodal points exhibit the non-Fermi-liquid features, which are more pronounced in the nodal region.


\vspace{-0.5cm}

\subsubsection{\label{sec:citeref}Spectral functions}




\begin{figure}[b!]
    \centering

    \includegraphics[width=0.49\linewidth]{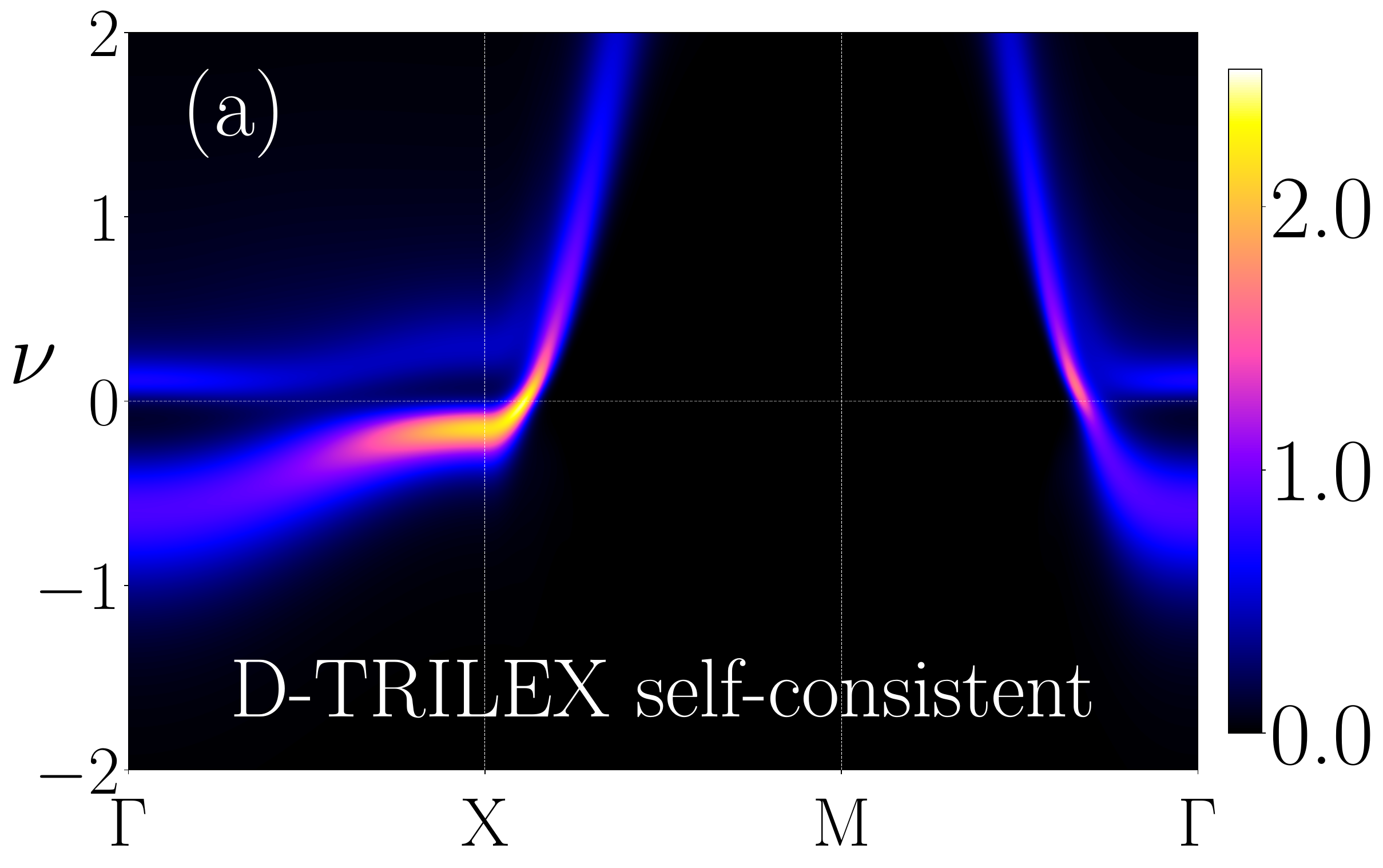}
\includegraphics[width=0.49\linewidth]{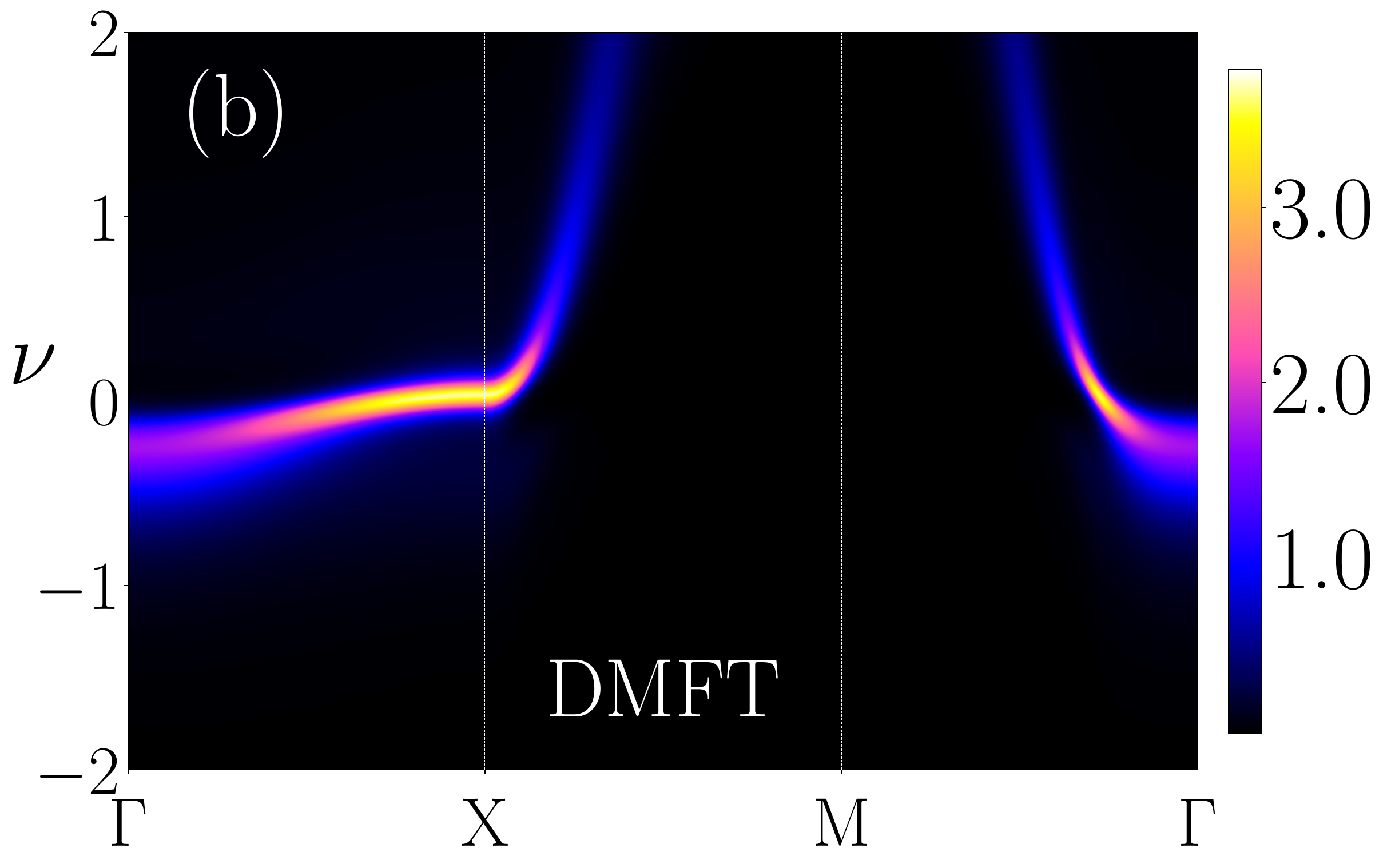}

    \includegraphics[width=0.49\linewidth]{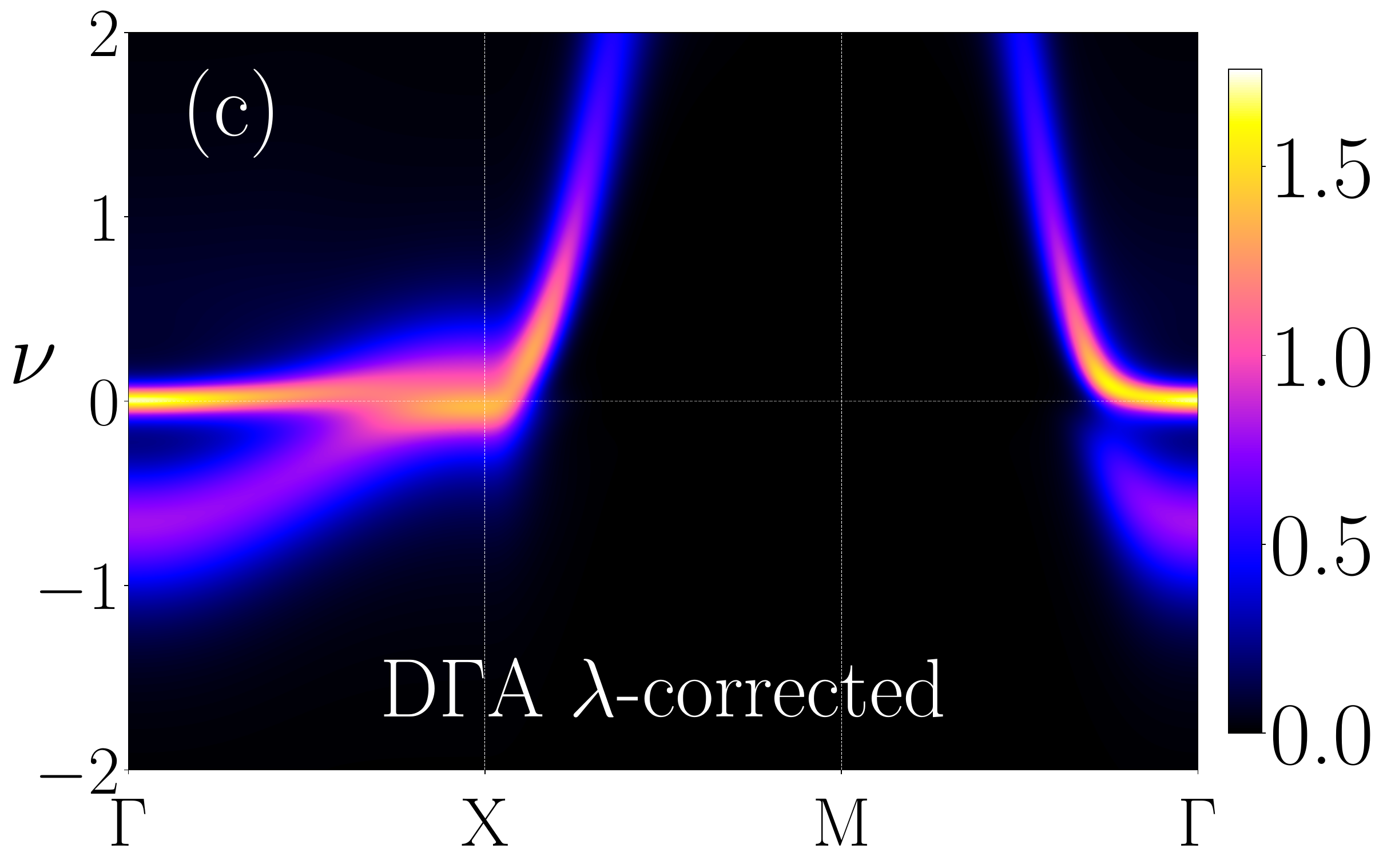}
\includegraphics[width=0.49\linewidth]{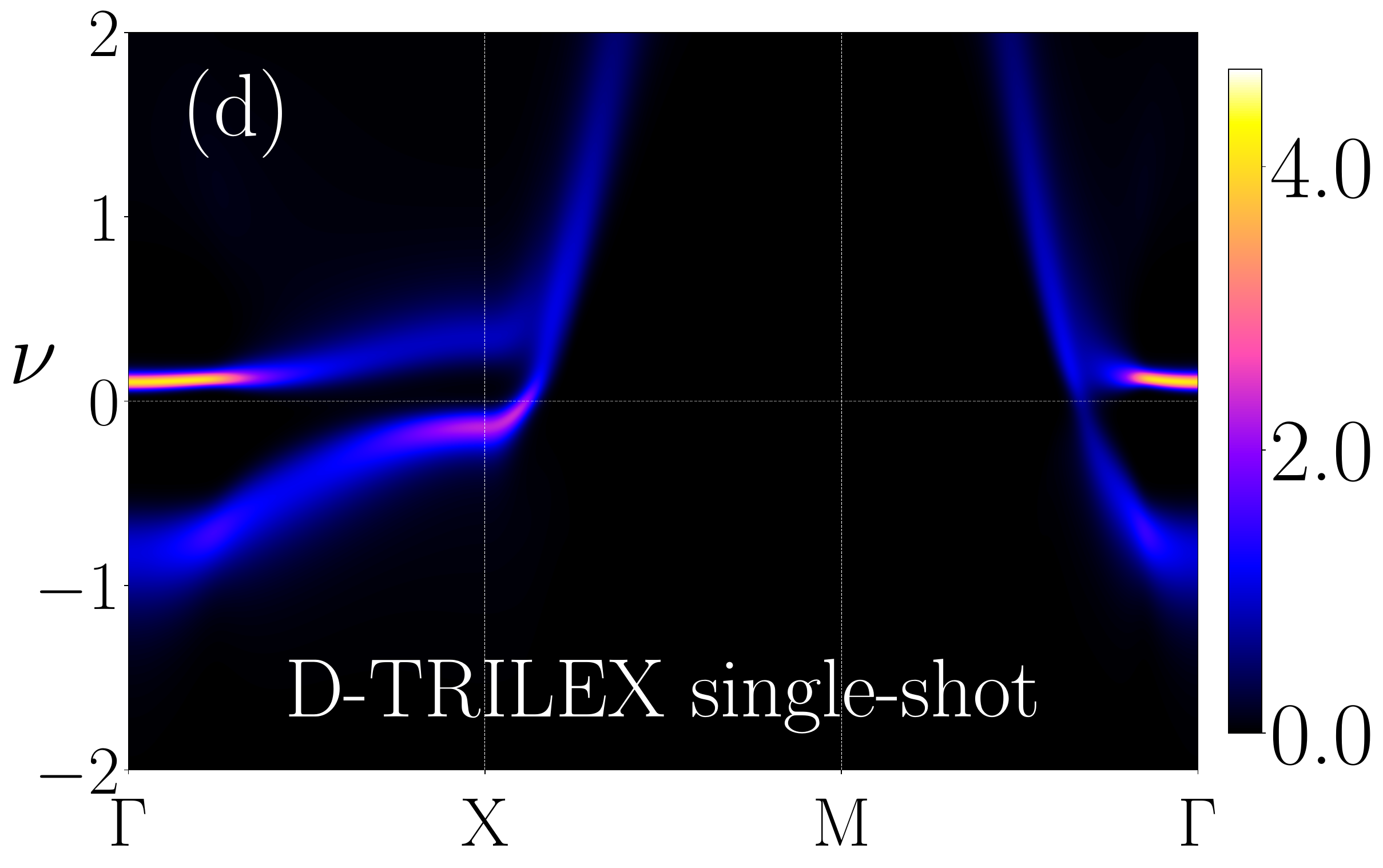}

    \includegraphics[width=0.49\linewidth]{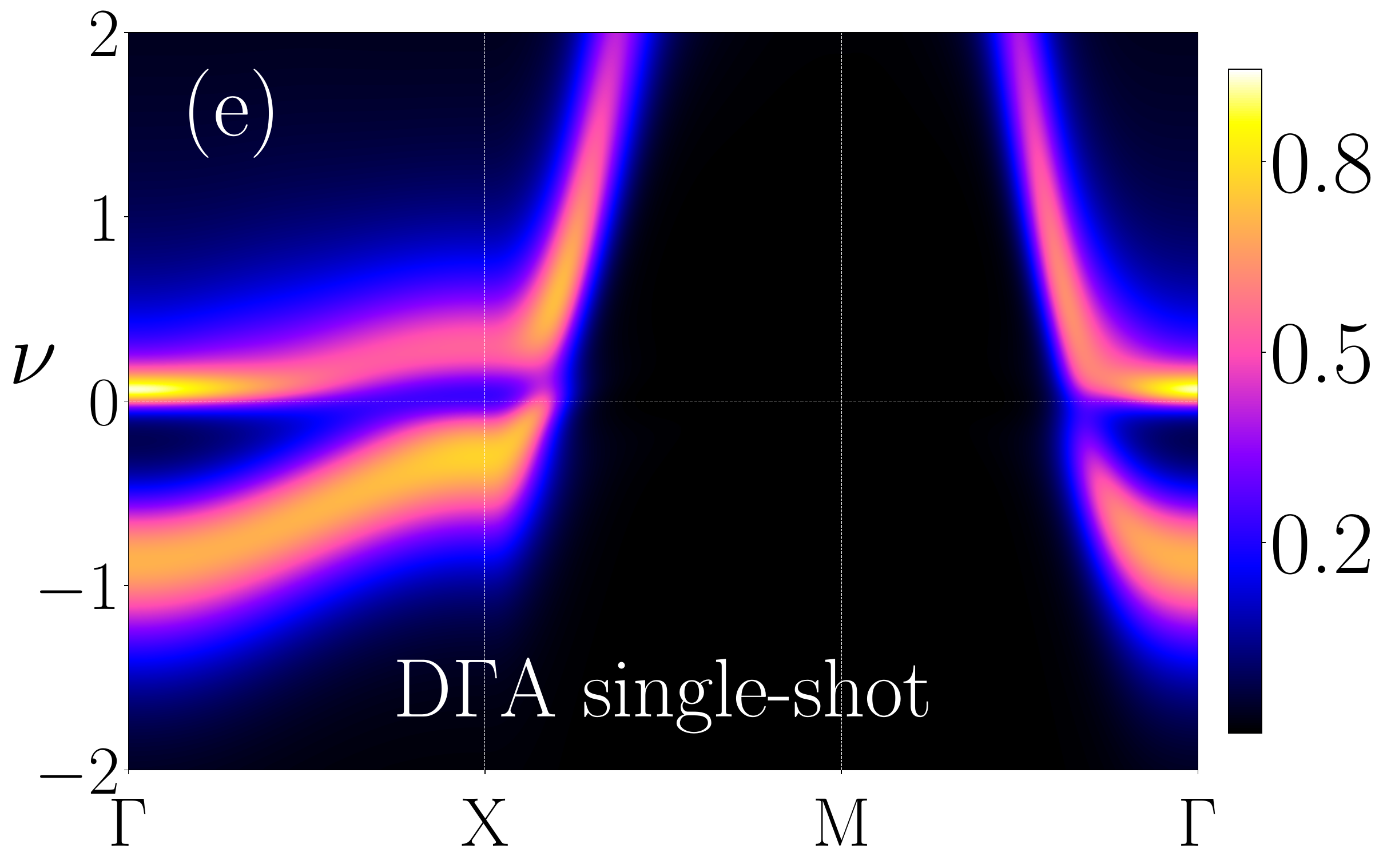}
    \includegraphics[width=0.49\linewidth]{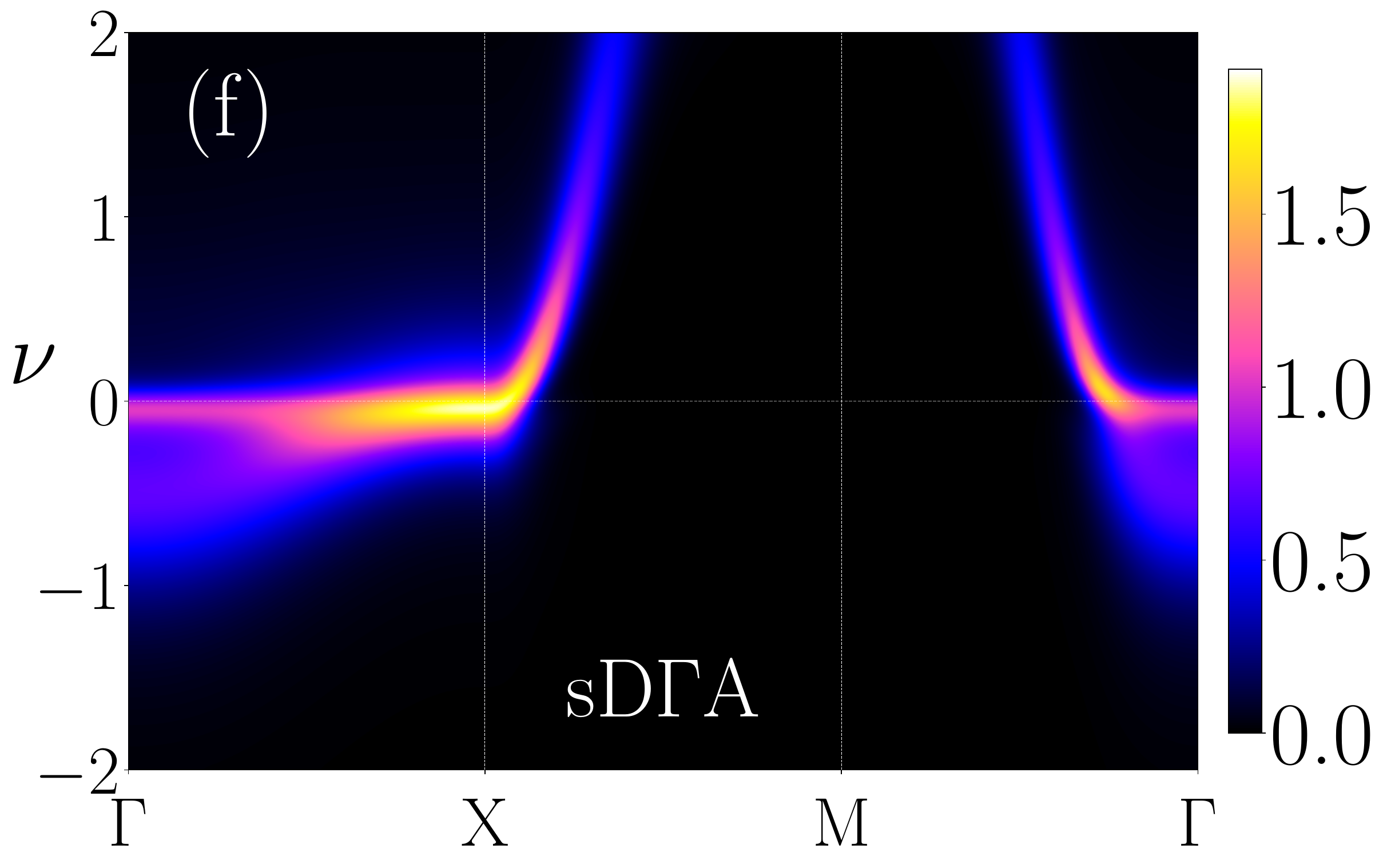}

    \caption{Momentum-resolved spectral functions for $T = 0.05$, $n = 0.43$. Panels (a–f): (a) \mbox{D-TRILEX}; (b) DMFT; (c) D$\Gamma$A with $\lambda$-correction;  (d) \mbox{D-TRILEX} single-shot; (e) D$\Gamma$A without $\lambda$ correction; (f) self-consistent D$\Gamma$A.}
    \label{fig:spect_GXMG}
\end{figure}

In Fig.~\ref{fig:spect_GXMG} we show spectral functions along the $\Gamma$--X--M--$\Gamma$ path in the Brillouin zone. 
In agreement with previous considerations~\cite{KK,KTr}, the non-local self-energy corrections within \mbox{D-TRILEX} lead to a characteristic splitting of the quasiparticle band near the Fermi level along the $\Gamma$--X and M--$\Gamma$ high-symmetry directions (see Fig.~\ref{fig:spect_GXMG}\,a), compared to the DMFT results (Fig.~\ref{fig:spect_GXMG}\,b). 
This behavior of the spectral function in \mbox{D-TRILEX} is induced by strong non-local correlations in the vicinity of the van Hove singularity. 
In particular, in contrast to the $\lambda$-corrected D$\Gamma$A \cite{QS4} (see also Fig. \ref{fig:spect_GXMG}\,c), where the splitting is mainly concentrated in the vicinity of $\Gamma$ point, we find only weakly momentum-dependent splitting in \mbox{D-TRILEX}. 

Also, in contrast to D$\Gamma$A, \mbox{D-TRILEX} yields more intensive lower (dispersive) band, while higher (flatter) band receives much smaller spectral weight and also possesses some dispersion. The latter band remains {\it above} the Fermi level and, therefore, does not lead to additional sheets of the Fermi surface (see below). Most importantly, both bands remain dispersive in \mbox{D-TRILEX} in contrast to the $\lambda$-corrected D$\Gamma$A approach.
Although a scenario of strong flattening of the electronic spectrum was proposed earlier for cuprates \cite{Khodel,Dzyaloshinskii,Khodel,RobvH}, in the considered case, perfectly flat electronic dispersion may occur as an  artifact of the $\lambda$-corrected D$\Gamma$A approach, which was also applied previously for this problem in Ref.~\cite{QS4}. The latter approach  somewhat suppresses the strength of the ferromagnetic correlations, yielding quasiparticle behavior of the upper band near the $\Gamma$ point, which in turn results in the flattening of the electronic spectrum.

To understand the origin of this difference between $\lambda$-corrected D$\Gamma$A and \mbox{D-TRILEX}, we also compare their results with the ``single shot'' non-self-consistent results, where the self-energy is evaluated with Green's functions, containing only local self-energy: the first iteration of \mbox{D-TRILEX} (Fig.~\ref{fig:spect_GXMG}d), as well 
D$\Gamma$A without $\lambda$-correction (Fig.~\ref{fig:spect_GXMG}e). One can see that while the first iteration of the \mbox{D-TRILEX} approach also yields stronger splitting at the $\Gamma$ point, the splitting is much less momentum dependent than in the $\lambda$-corrected D$\Gamma$A. Moreover, a similar weak momentum dependence of the splitting is observed in the D$\Gamma$A approach without the $\lambda$-correction (Fig.~\ref{fig:spect_GXMG}e). From this, we conclude that the flat band and strong momentum dependence of the splitting are likely the artifacts of $\lambda$-corrected D$\Gamma$A. For comparison, we also show the result of the self-consistent D$\Gamma$A ~\cite{SCDGA}. In this case, the spectral weight is concentrated near the X point, forming an extended van Hove singularity, 
which is similar to that in \mbox{D-TRILEX}, however, with suppressed splitting of the electronic spectrum. The intensity of the flat band is also suppressed in the self-consistent D$\Gamma$A, in comparison to
the $\lambda$-corrected D$\Gamma$A, which has intensity 
maximum at the $\Gamma$ point.

\begin{figure}[t!]
    \centering
    \includegraphics[width=0.475\linewidth]{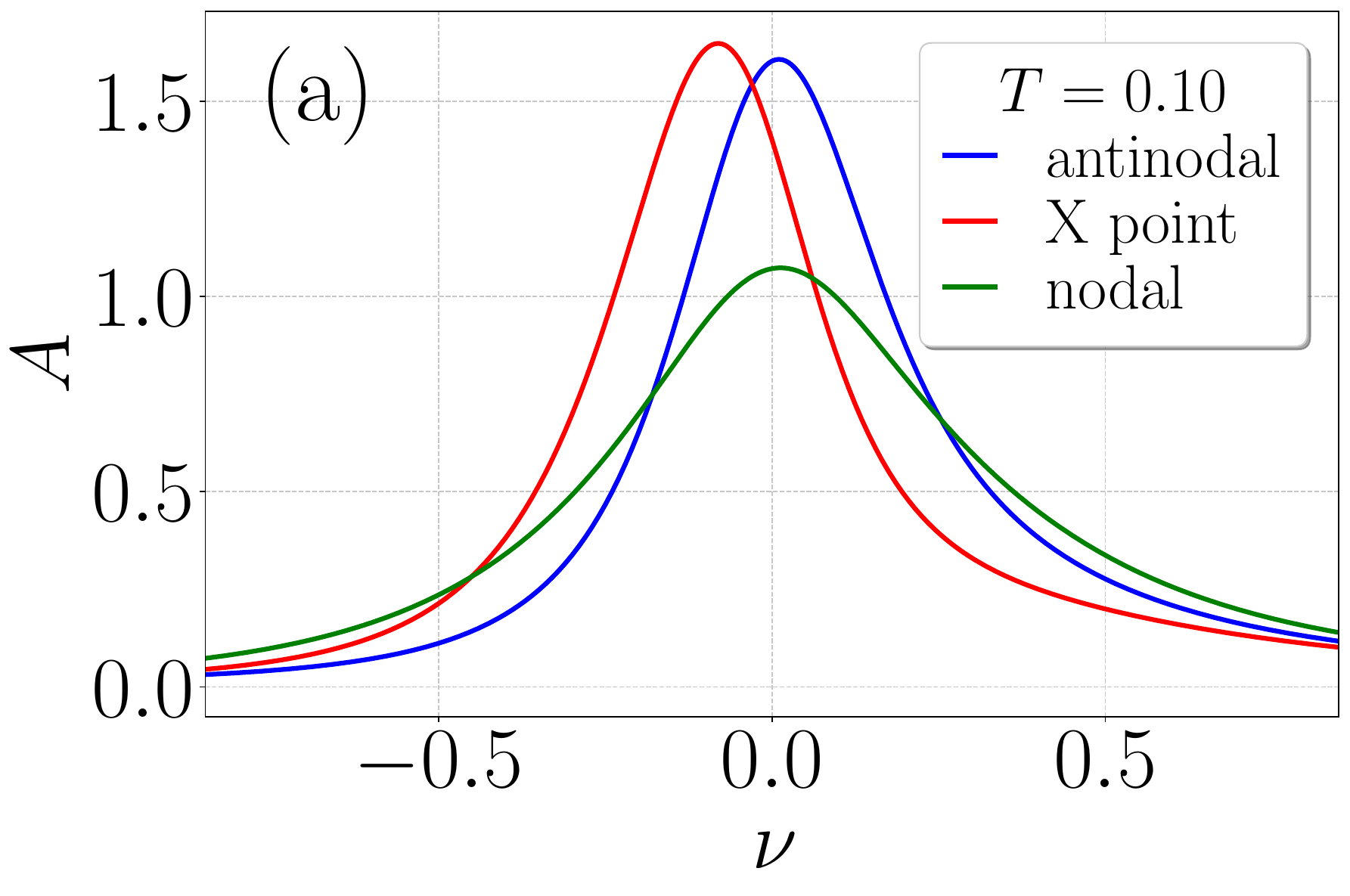}
    \includegraphics[width=0.475\linewidth]{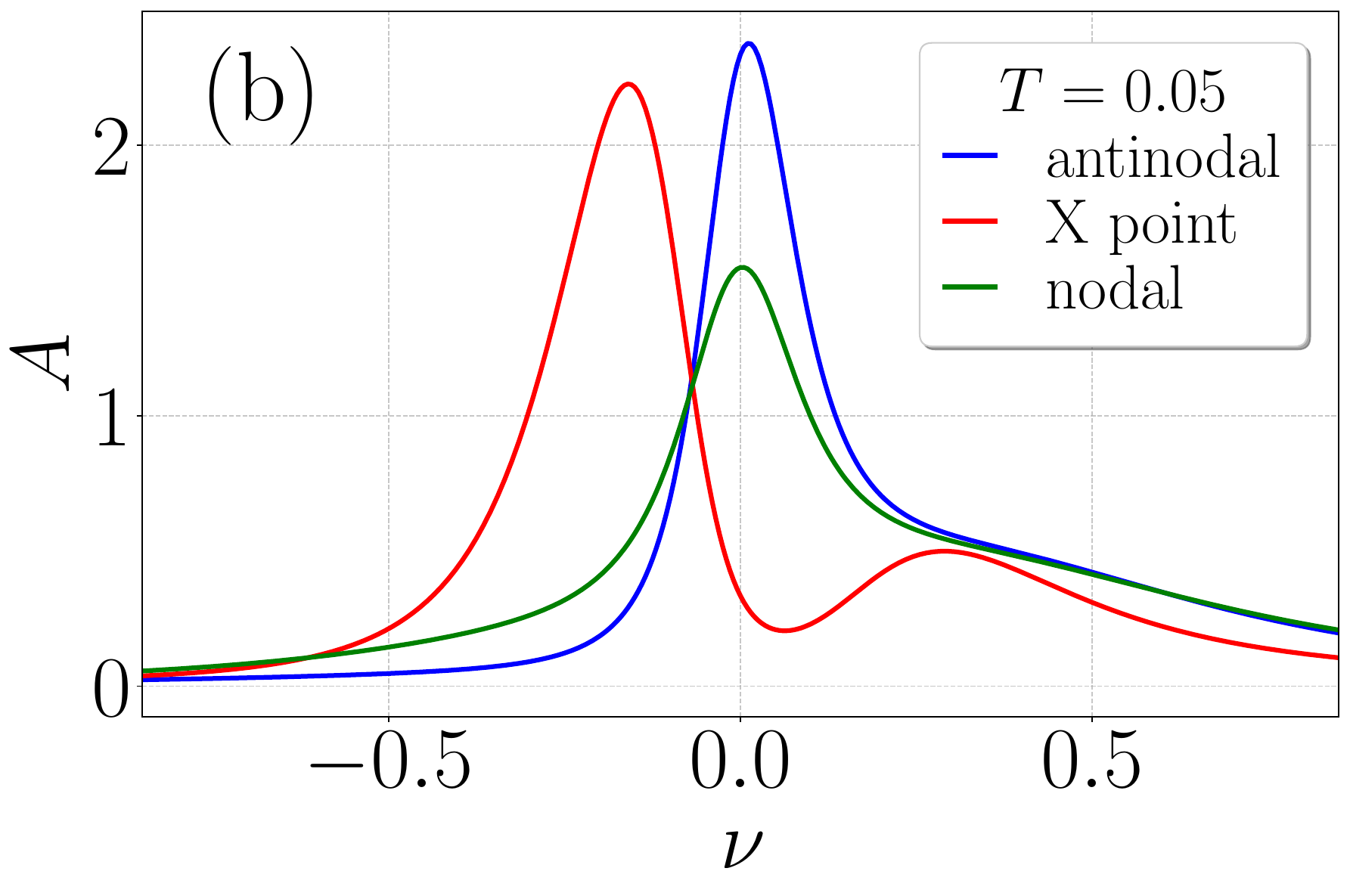}
    \caption{Spectral functions in \mbox{D-TRILEX} at $n = 0.43$ at the X point (red line), nodal and antinodal points (green and blue lines) (a) $T = 0.1$, (b) $T = 0.05$
} 
    \label{fig:SpecT}
\end{figure}

The electronic spectral functions at the antinodal, nodal, and X points calculated at ${T = 0.1}$ and ${T = 0.05}$ are shown in Fig.~\ref{fig:SpecT}. 
It can be seen that, {in the presence of the ferromagnetic fluctuations}, the electronic excitations at the antinodal point are more coherent than those at the nodal point,
{which is opposite to the antiferromagnetic case}. The ratio of spectral weights at the nodal and antinodal points remains the same for both temperatures. As the temperature decreases, the spectral weight becomes suppressed at the X point; the quasiparticle damping at the nodal and antinodal points decreases, causing the peaks of the spectral functions to become narrower.

\begin{figure}[t!]
    \centering
    \includegraphics[width=0.95\linewidth]{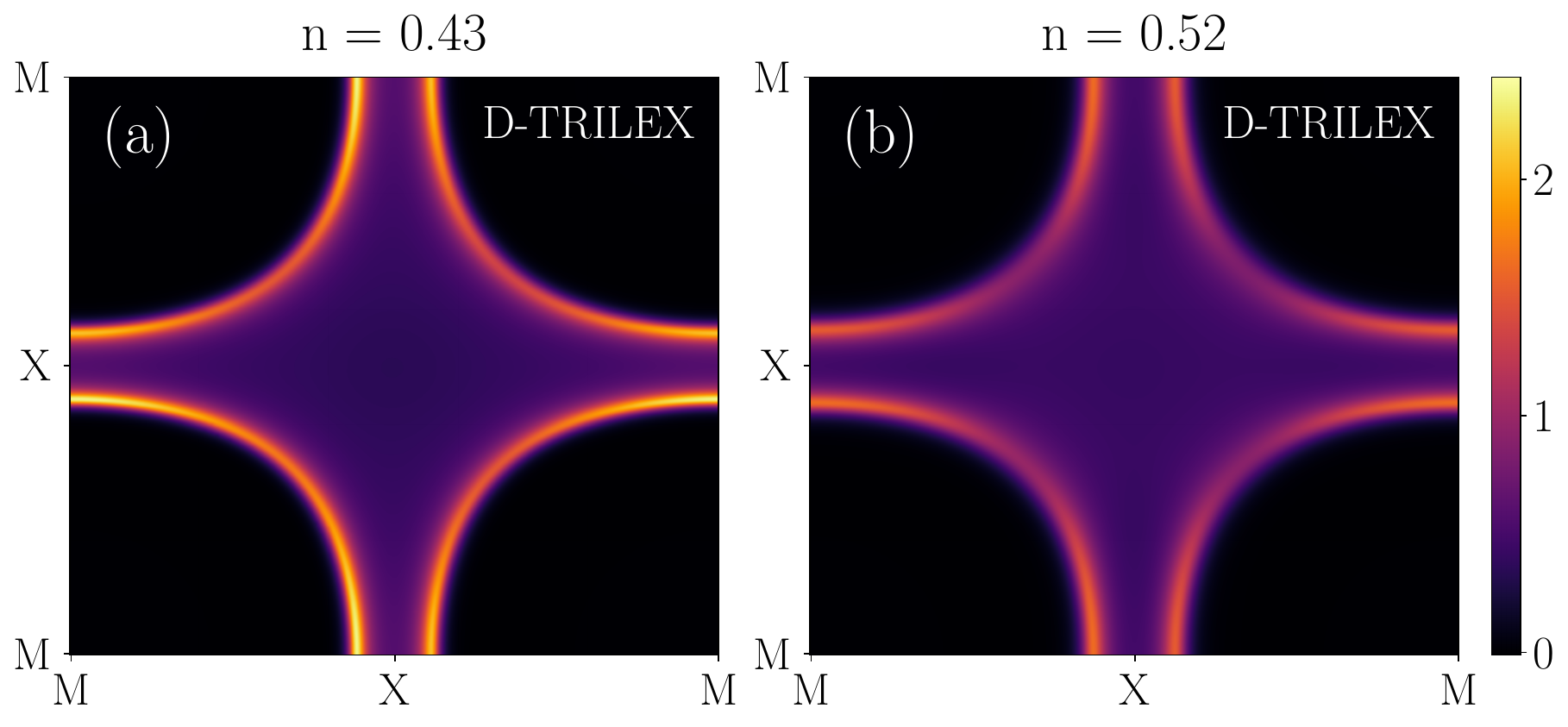}

    \includegraphics[width=0.95\linewidth]{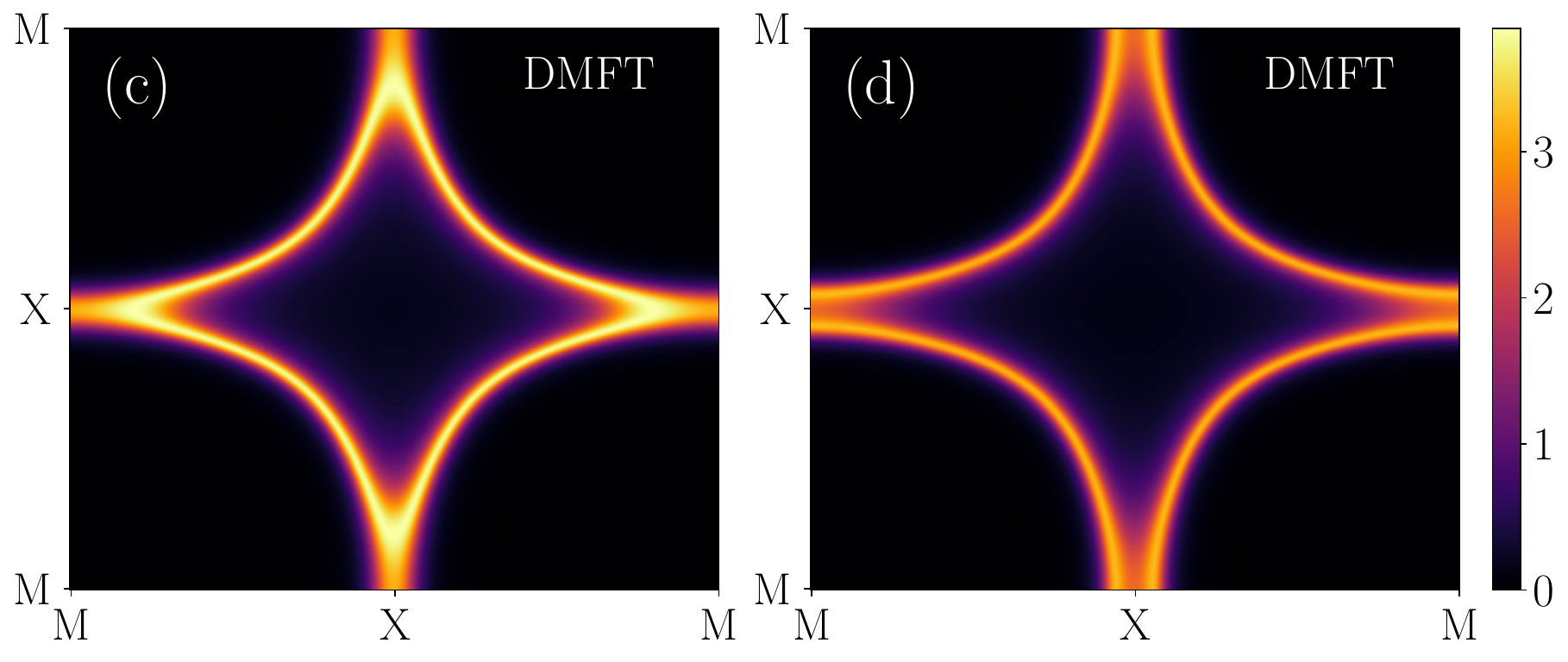}
    
\includegraphics[width=0.95\linewidth]{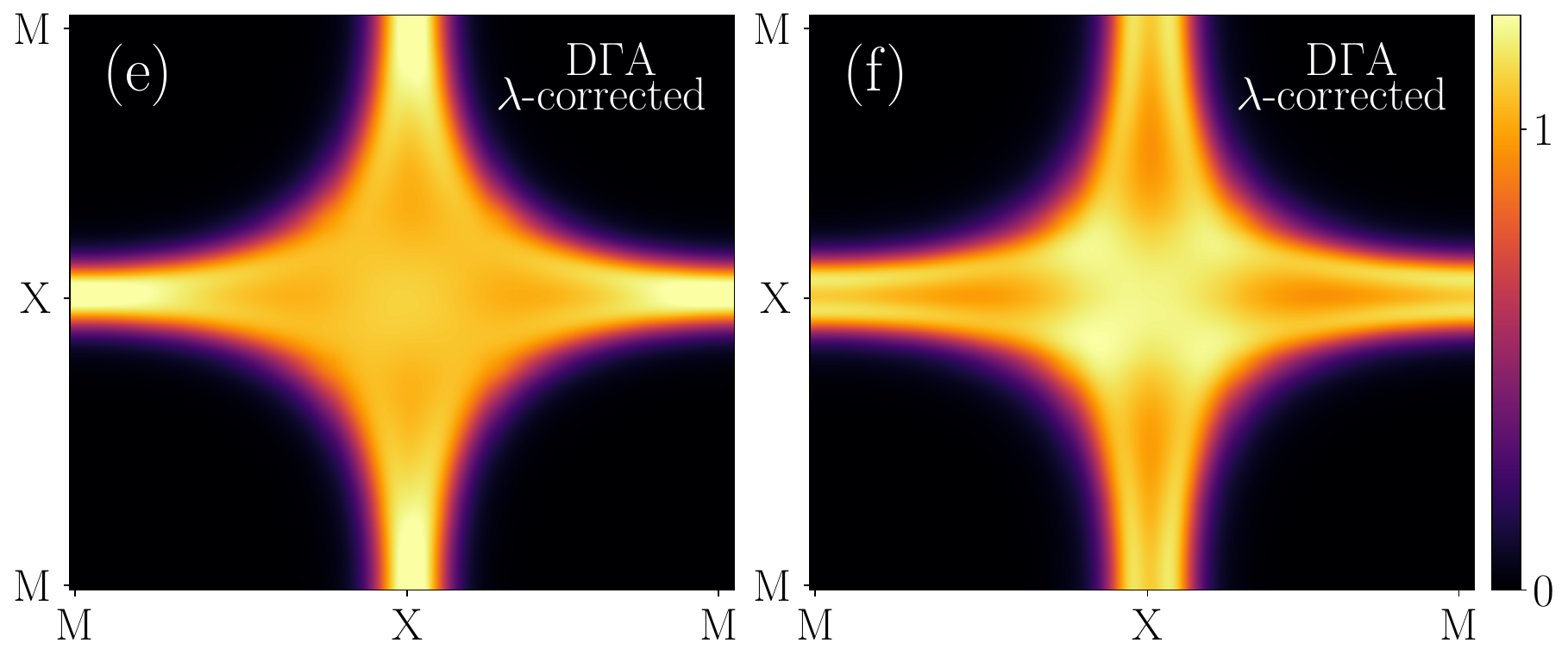}

    \caption{Spectral functions at $\nu = 0$ in the Brillouin zone for $T = 0.05$ in \mbox{D-TRILEX} approach (a,b), compared to DMFT (c,d) and $\lambda$-corrected D$\Gamma$A (e,f) at $n = 0.43$ (a,c,e) and $n = 0.52$ (b,d,f).}
    \label{fig:spect_FS}
\end{figure}

The resulting Fermi surfaces in \mbox{D-TRILEX} are shown in Fig.~\ref{fig:spect_FS}\,a,\,b. 
Despite the mentioned reshaping of the dispersion, the Fermi surface remains unsplit since one of the two bands does not cross the Fermi level. Moreover, non-local corrections lead to a spectral weight suppression  near the nodal point. In particular, the redistribution of spectral weight and the expansion of the Fermi surface demonstrate that non-local fluctuations can strongly modify the low-energy electronic structure while preserving the overall Fermi surface topology. As the filling increases from ${n = 0.43}$ to ${n = 0.52}$ the spectral function broadens, while its angular dependence remains preserved. The obtained Fermi surfaces violate Luttinger theorem, which is related to the fact that the obtained excitations carry only part of the spectral weight, while another part is split into an upper weakly dispersive band above the Fermi level (see Fig.~\ref{fig:spect_GXMG}). Therefore, it is natural that the Fermi surface expands to compensate for this ``missing'' spectral weight. 
These features are absent in the local DMFT solution (Figs.~\ref{fig:spect_FS}\,c,\,d), where the spectral function retains a single quasiparticle band with only moderate broadening. The comparison highlights that the inclusion of non-local self-energy effects in \mbox{D-TRILEX} is crucial to capture the qualitative changes in momentum-resolved spectra, in agreement with the stronger quasiparticle damping discussed above.

In Figs.~\ref{fig:spect_FS}\,e,\,f we also compare the obtained Fermi surfaces to the $\lambda$-corrected D$\Gamma$A result. Despite the similarity in the shape of the Fermi surface with D-TRILEX, in  D$\Gamma$A, the Fermi surface appears to be strongly broadened and not well defined, which is related to the too strong flatness of the electronic dispersion in this approach.  

\section{Conclusion}


We investigated the non-Fermi-liquid properties of the Hubbard model in the vicinity of the van Hove singularity (vHS) with strong ferromagnetic fluctuations within DMFT and \mbox{D-TRILEX}. The latter approach considers the effect of the non-local self-energy corrections self-consistently.  The treatment of the non-local correlations enhances non-Fermi-liquid properties: while in DMFT we find the quasiparticle damping $\gamma\sim T^2$ at low temperatures, in \mbox{D-TRILEX} in the vicinity of the van Hove singularity, we obtain $\gamma\sim {\rm const}(T)$ 
(we find $\gamma$ even slightly increases with decreasing $T$ closer to the $T^*$). 
At the nodal and antinodal points of the Fermi surface, the damping decreases with temperature, but the Fermi liquid behavior is not restored. 

The spectral functions in \mbox{D-TRILEX} exhibit split bands, one of which lies  above the Fermi level, leading to an expansion of the Fermi surface to conserve the number of particles. Spectral functions and self-energies show a pronounced nodal–antinodal dichotomy, with higher spectral weight at the nodal point, contrary to the antiferromagnetic scenario.

We also analyzed the role of self-consistency of non-local quantities through a comparison of the first and last iterations of D-TRILEX and D$\Gamma$A. The latter approach in the most widely used $\lambda$-corrected version leads to a flat upper band. By comparing this result to one-shot approaches, we conclude that this is likely an artifact of the $\lambda$-correction. This drawback is partly improved in the self-consistent D$\Gamma$A approach; however, it leads to a smaller splitting of the electronic spectrum. 


These results demonstrate that the presence of the van Hove singularity drives strong ferromagnetic fluctuations, and that non-local self-energies play a decisive role in shaping the spectral properties, particularly the splitting of the band and the nodal–antinodal dichotomy. Importantly, we find that the features of this splitting and the dichotomy are  method-dependent, while self-consistency and considering local vertex corrections in D-TRILEX are important for accurately capturing these effects.

Further studies could extend this analysis to multi-orbital systems within the \mbox{D-TRILEX} framework~\cite{DTRILEX3} and density-density interaction, {which can be efficiently handled by the segment CT-QMC hybridization expansion \mbox{iQIST} solver~\cite{iQIST}}. 
Such extensions would provide a way to investigate itinerant ferromagnets such as nickel or iron, where the interplay between orbital degrees of freedom and non-local spin fluctuations also plays an important role. Combining \mbox{D-TRILEX} with density functional theory (DFT)+DMFT input would provide a realistic route to describe correlated magnetic metals with strong momentum-dependent electronic self-energies.

\vspace{-0.1cm}

\section*{Acknowledgements}
Performing the DMFT calculations (I.S.D., A.A.K.) was supported by the Russian Science Foundation (project 24-12-00186). 
Development of the D$\Gamma$A and self-consistent D$\Gamma$A programs was carried out within the framework of the state assignment of the Ministry of Science and Higher Education of the Russian Federation for the IMP UB RAS. The calculations were performed on the cluster of the Laboratory of Material Computer Design of MIPT.

\renewcommand\theequation{A\arabic{equation}}
\renewcommand\thefigure{A\arabic{figure}}
\setcounter{equation}{0}
\setcounter{figure}{0}
\vspace{-0.2cm}
\subsection*{Appendix}

\begin{figure}[h!]
    \centering
    \includegraphics[width=0.46\linewidth]{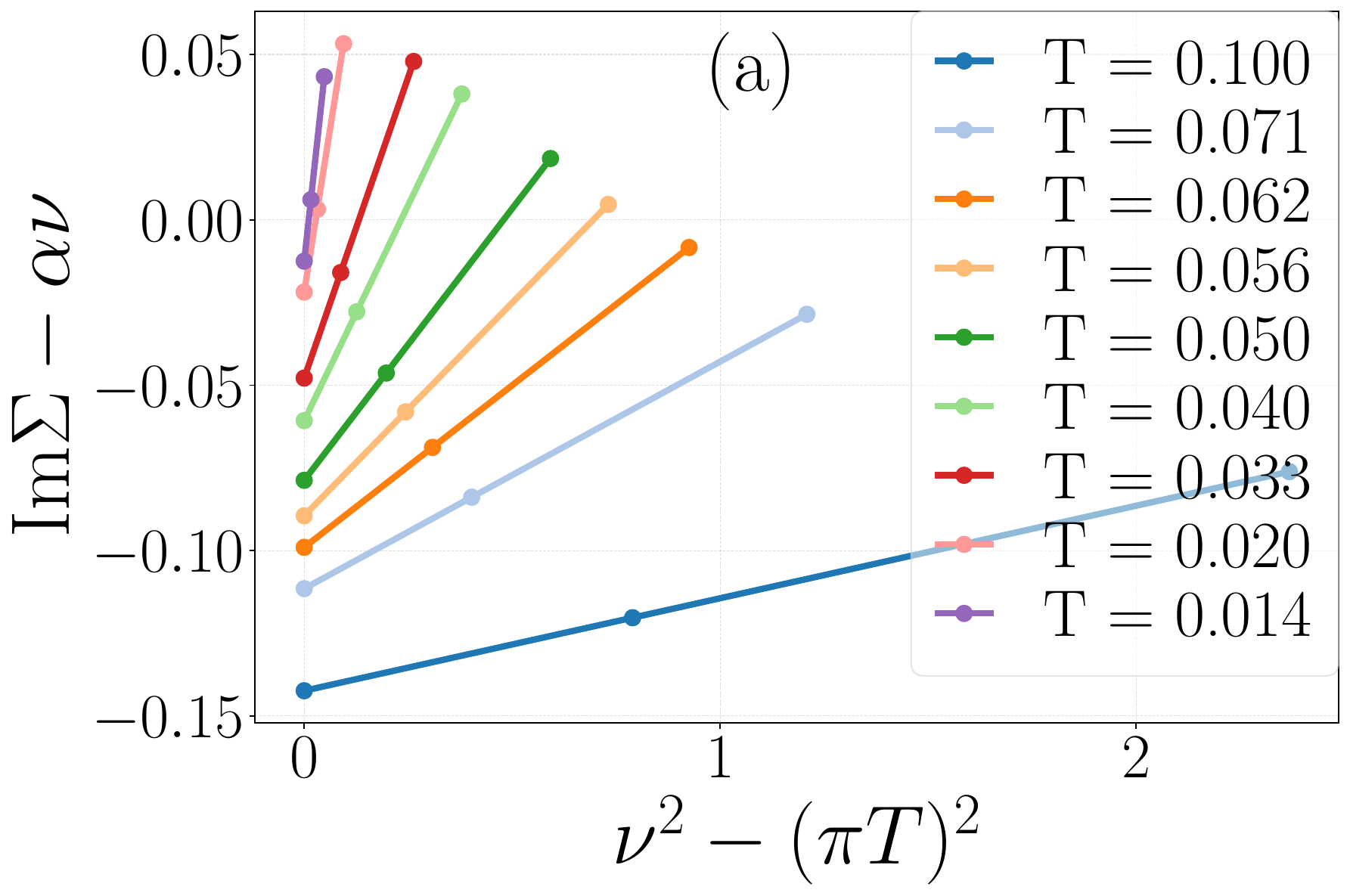}
    \includegraphics[width=0.46\linewidth]{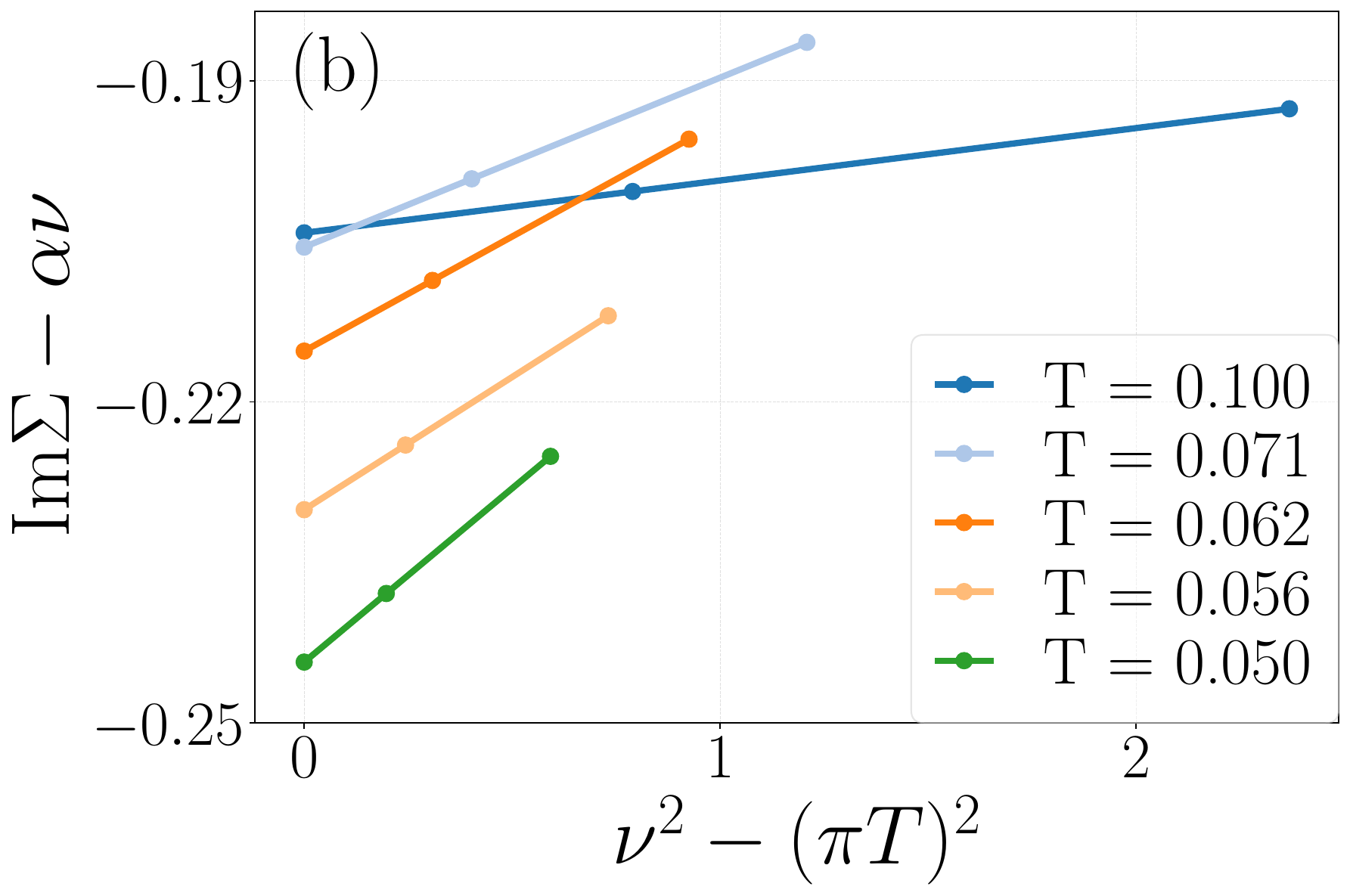}
    \includegraphics[width=0.46\linewidth]{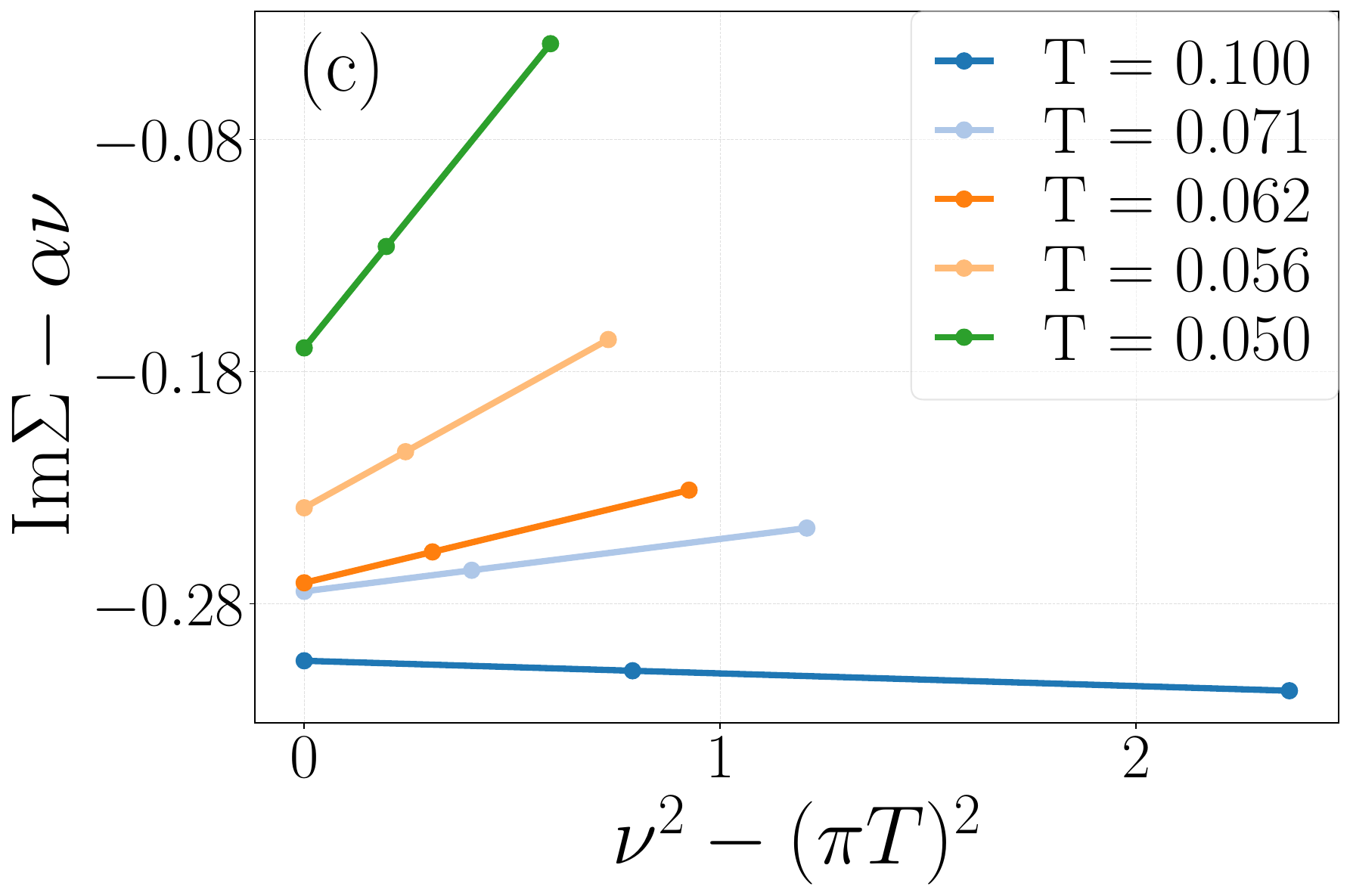}
    \includegraphics[width=0.46\linewidth]{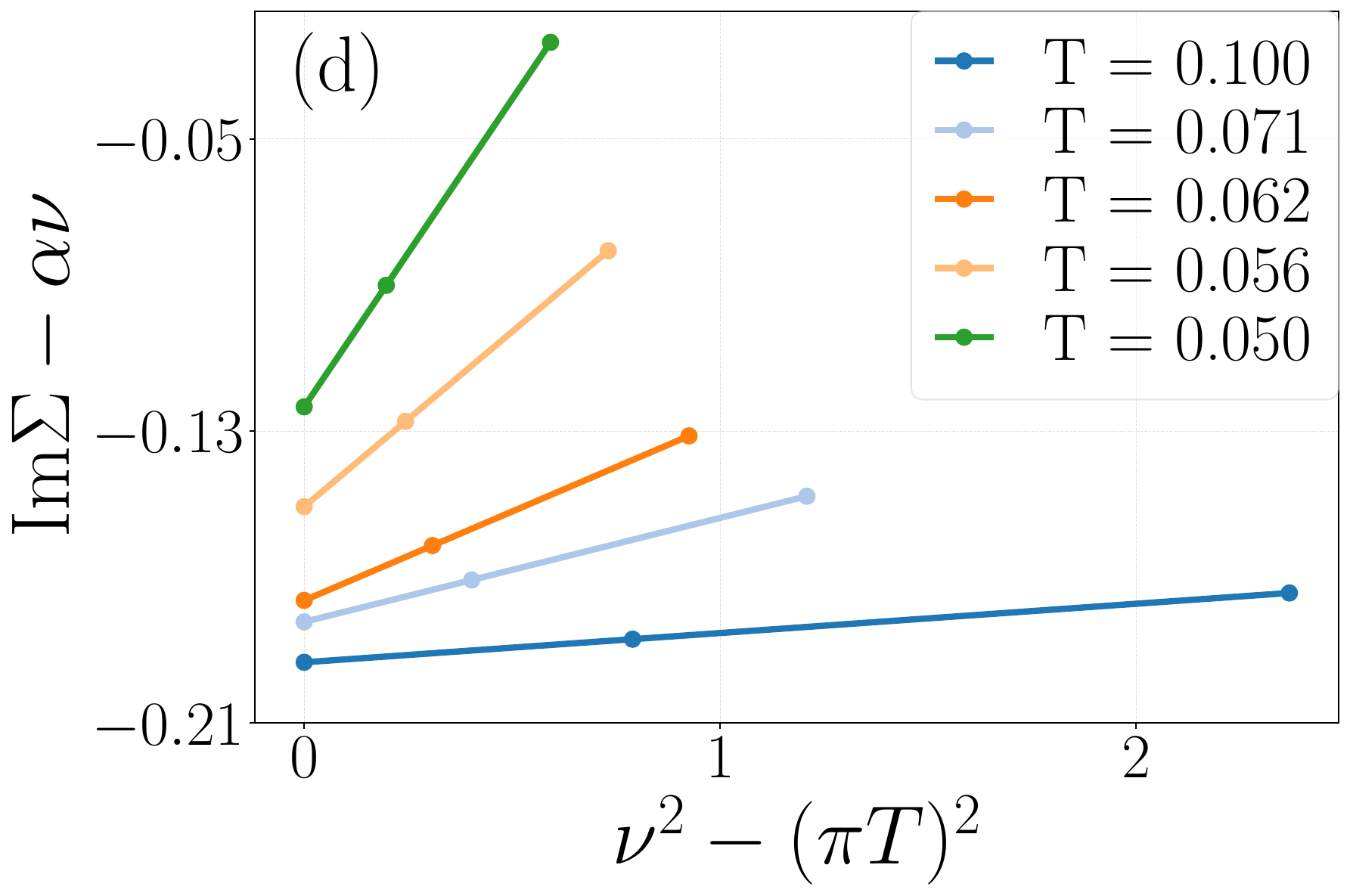}

    \includegraphics[width=0.46\linewidth]{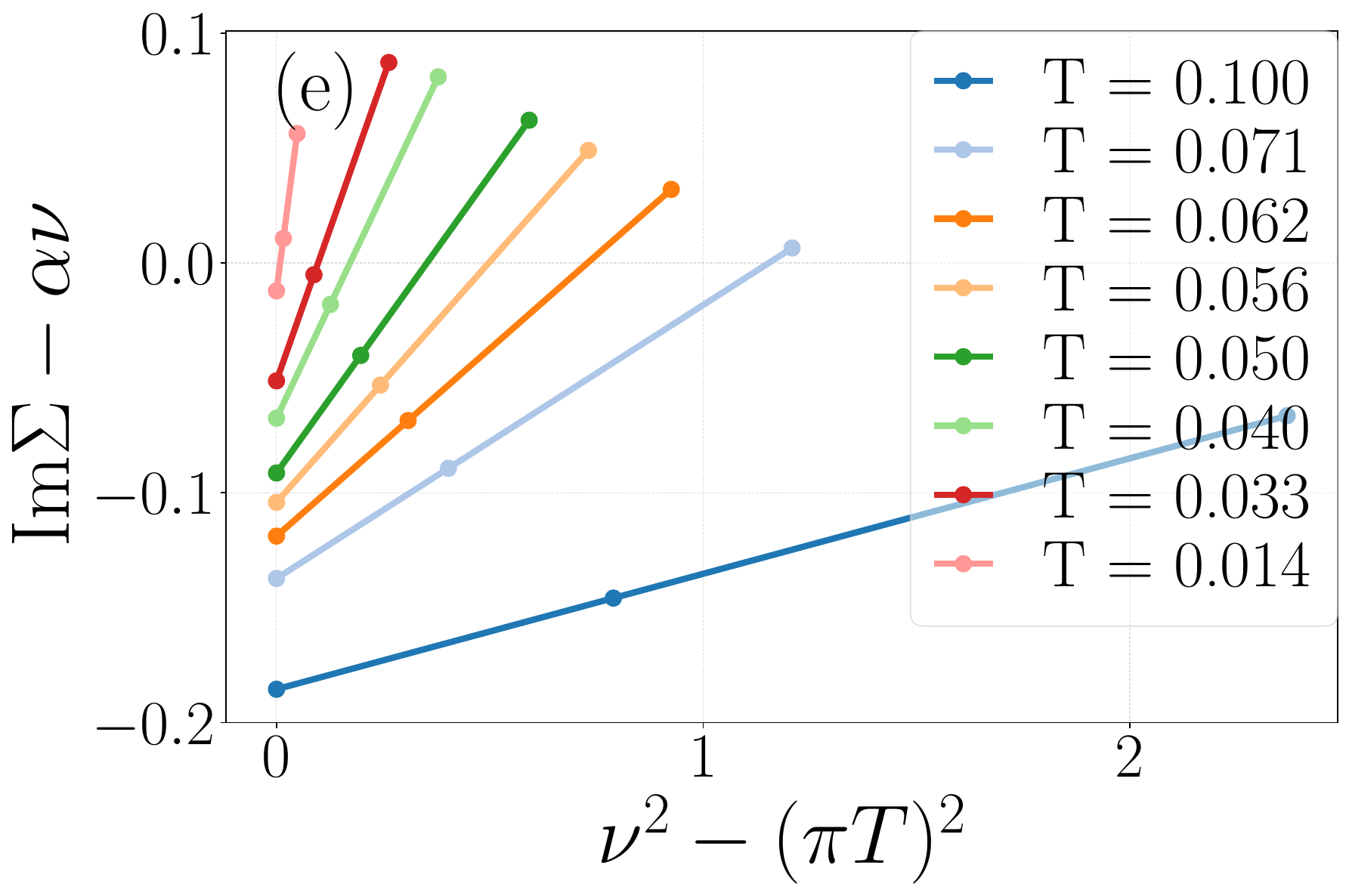}
    \includegraphics[width=0.46\linewidth]{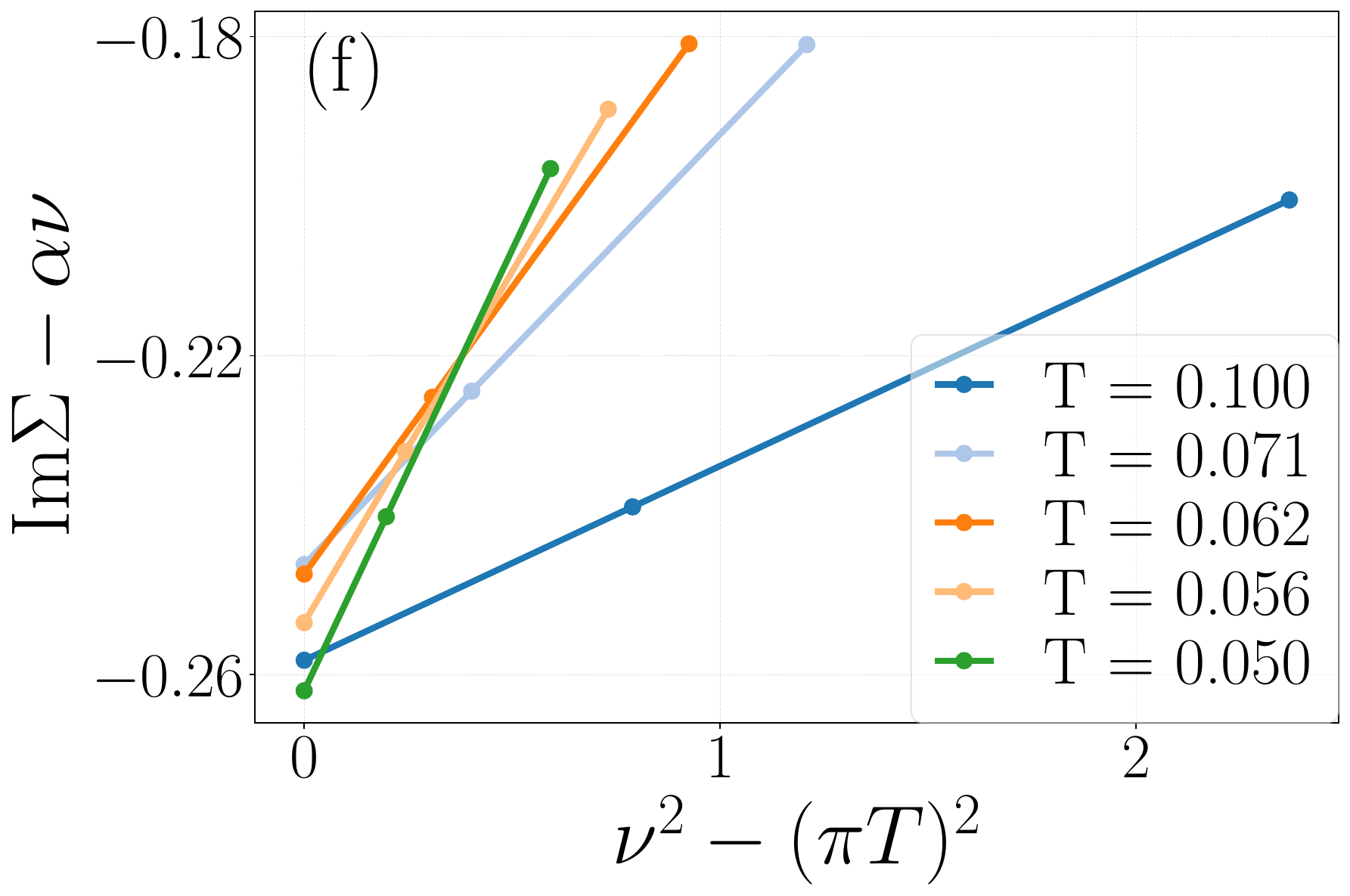}
    \includegraphics[width=0.46\linewidth]{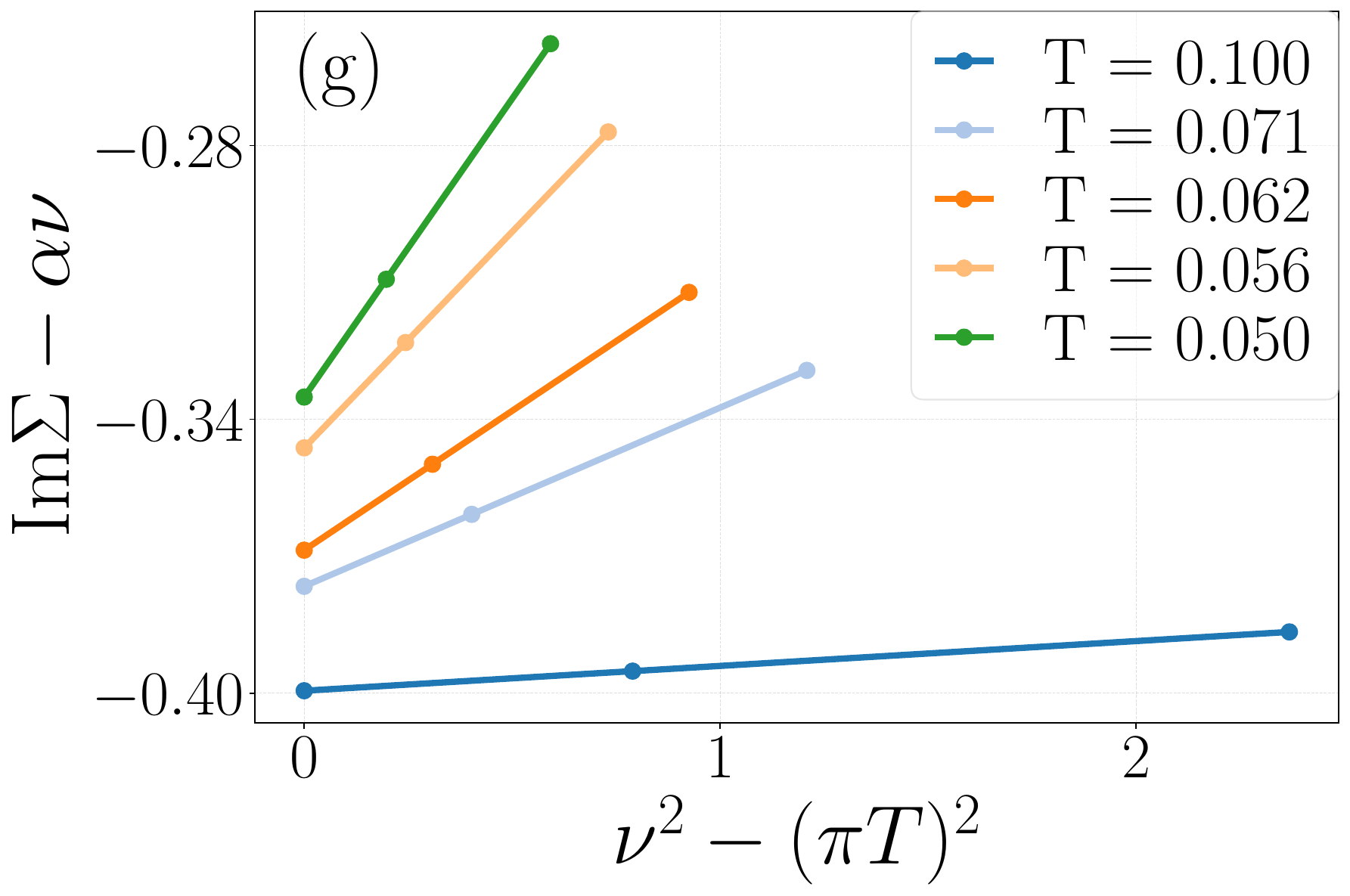}
    \includegraphics[width=0.46\linewidth]{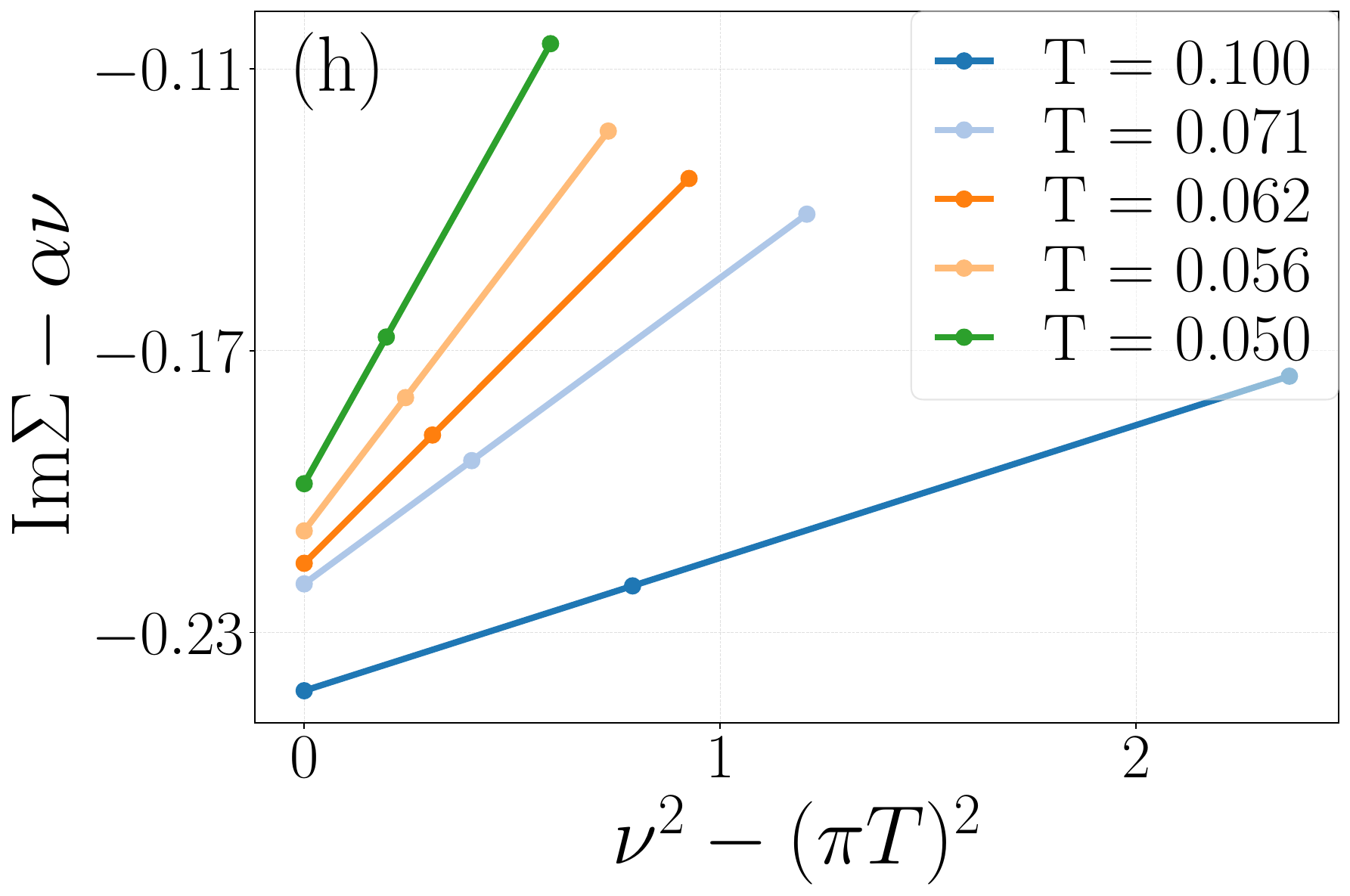}
    \caption{Im$\Sigma (\nu) - \alpha\nu$, as a function of $\nu^2 - (\pi T)^2$, which is linear in Fermi-Liquid theory, at  $n=0.43$ (a-d) and $n=0.52$ (e-f) in DMFT (a,e), and \mbox{D-TRILEX}  X point (b,f), nodal point (c,g),  and antinodal point (d,h).}
    \label{fig:nu2T2}
\end{figure}

In this Appendix we present the dependencies of the self-energies $\text{Im}\Sigma (\nu) - \alpha\nu$ on $\nu^2 - (\pi T)^2$ (see main text). The obtained dependencies for ${n=0.43}$ and ${n=0.52}$ are shown in Fig.~\ref{fig:nu2T2} and show Fermi-liquid-like behavior of self-energies in DMFT at low temperatures and non-Fermi-liquid behavior in \mbox{D-TRILEX}.

\vspace{0.2cm}

\newpage

\begin{thebibliography}{99}
\bibitem{Chubukov} A. V. Chubukov, D. Pines, and B. P. Stojkovi\'c, Temperature crossovers in cuprates, \href{https://dx.doi.org/10.1088/0953-8984/8/48/021}{J. Phys.: Condens. Matter {\bf 8}, 10017 (1996)}.

\bibitem{Pines1} B. P. Stojkovi\'c and D. Pines, Theory of the optical conductivity in the cuprate superconductors, \href{https://doi.org/10.1103/PhysRevB.56.11931}{Phys. Rev. B {\bf 56}, 11931 (1997)}.

\bibitem{Pines} D. Pines, Nearly antiferromagnetic Fermi liquids: a progress report, \href{https://doi.org/10.1007/s002570050346}{Z. Phys. B {\bf 103}, 129 (1997)}. 


\bibitem{Tremblay}  Y. M. Vilk and A.-M. S. Tremblay, Non-perturbative many-body approach to the Hubbard model and single-particle pseudogap, \href{https://doi.org/10.1051/jp1%3A1997135}{J. Phys. I France {\bf 7}, 1309 (1997)}.

\bibitem{SPS} J. Schmalian, D. Pines, and B. Stojkovi\'c, Microscopic theory of weak pseudogap behavior in the underdoped cuprate superconductors: General theory and quasiparticle properties, \href{https://doi.org/10.1103/PhysRevB.60.667}{Phys. Rev. B {\bf 60}, 667 (1999)}.

\bibitem{PG0} O. Gunnarsson, T. Schäfer, J. P. F. LeBlanc, E. Gull, J. Merino, G. Sangiovanni, G. Rohringer, and A. Toschi, Fluctuation diagnostics of the electron self-energy: Origin of the pseudogap physics, \href{https://doi.org/10.1103/PhysRevLett.114.236402}{Phys. Rev. Lett. {\bf 114}, 236402 (2015)}.

\bibitem{PG1} F. Krien, P. Worm, P. Chalupa, A. Toschi, and K. Held, Explaining the pseudogap through damping and antidamping on the Fermi surface by imaginary spin scattering, \href{https://doi.org/10.1038/s42005-022-01117-5}{Commun. Phys. {\bf 5}, 336 (2022)}.

\bibitem{PRL2024} M. Chatzieleftheriou, S. Biermann, and E. A. Stepanov, Local and non-local electronic correlations at the metal-insulator transition in the two-dimensional Hubbard model, \href{https://doi.org/10.1103/PhysRevLett.132.236504}{Phys. Rev. Lett. {\bf 132}, 236504 (2024)}.

\bibitem{PG2} Y. Yu, S. Iskakov, E. Gull, K. Held, F. Krien, Unambiguous fluctuation decomposition of the self-energy: pseudogap physics beyond spin fluctuations, \href{https://doi.org/10.1103/PhysRevLett.132.216501}{Phys. Rev. Lett. {\bf 132}, 216501 (2024)}.

\bibitem{PG3} J.-M. Lihm, Dominik Kiese, S.-S. B. Lee, F. B. Kugler, The finite-difference parquet method: Enhanced electron-paramagnon scattering opens a pseudogap, \href{https://doi.org/10.48550/arXiv.2505.20116}{ArXiv:2505.20116}.

\bibitem{KK} A. A. Katanin, A. P. Kampf, and V. Yu. Irkhin, Anomalous self-energy and Fermi surface quasi-splitting in the vicinity of a ferromagnetic instability, \href{https://doi.org/10.1103/PhysRevB.71.085105}{Phys. Rev. B {\bf 71}, 085105 (2005)}.



\bibitem{KTr} A. A. Katanin, Electronic self-energy and triplet pairing fluctuations in the vicinity of a ferromagnetic instability in 2D systems: the quasistatic approach, \href{https://doi.org/10.1103/PhysRevB.72.035111}{Phys. Rev. B {\bf 72}, 035111 (2005)}.


\bibitem{QS4} M. Kitatani, Y. Nomura, S. Sakai, and R. Arita, Luttinger surface and exchange splitting induced by ferromagnetic fluctuations, \href{https://doi.org/10.48550/arXiv.2509.21034}{ArXiv:2509.21034}.

\bibitem{QS3} Y. Nomura, S. Sakai, and R. Arita, Fermi Surface Expansion above Critical Temperature in a Hund
Ferromagnet, \href{https://doi.org/10.1103/PhysRevLett.128.206401}{Phys. Rev. Lett. {\bf 128}, 206401 (2022)}.

\bibitem{SrRu0} A. Liebsch and A. Lichtenstein, Photoemission Quasiparticle Spectra of Sr$_2$RuO$_4$, \href{https://doi.org/10.1103/PhysRevLett.84.1591}{Phys. Rev. Lett. {\bf 84}, 1591 (2000)}. 

\bibitem{SrRu1} V. B. Zabolotnyy, D. V. Evtushinsky, A. A. Kordyuk, T. K. Kim, E. Carleschi, B. P. Doyle, R. Fittipaldi, M. Cuoco, A. Vecchione, and S. V. Borisenko, Renormalized Band Structure of Sr$_2$RuO$_4$: A Quasiparticle Tight-Binding Approach, \href{https://doi.org/10.1016/j.elspec.2013.10.003}{J. Electron Spectrosc. Relat. Phenom. {\bf 191}, 48 (2013)}.

\bibitem{HundMetal} E. A. Stepanov, Y. Nomura, A. I. Lichtenstein, and S. Biermann, Orbital isotropy of magnetic fluctuations in correlated electron materials induced by Hund's exchange coupling, \href{https://doi.org/10.1103/PhysRevLett.127.207205}{Phys. Rev. Lett. {\bf 127}, 207205 (2021)}.

\bibitem{SrRu2} M. Chatzieleftheriou, A. N. Rudenko, Y. Sidis, S. Biermann, and E. A. Stepanov, Nature of momentum- and orbital-dependent magnetic fluctuations in ${\mathrm{Sr}}_{2}{\mathrm{RuO}}_{4}$, \href{https://doi.org/10.1103/ts6y-zb6m}{Phys. Rev. B {\bf 112}, 195118 (2025)}.

\bibitem{SrRuFM} T. Imai, A. W. Hunt, K. R. Thurber, and F. C. Chou, $^{17}$O NMR Evidence for Orbital Dependent Ferromagnetic Correlations in Sr$_2$RuO$_4$, \href{https://doi.org/10.1103/PhysRevLett.81.3006}{Phys. Rev. Lett. {\bf 81}, 3006 (1998)}.
\bibitem{Steffens} P. Steffens, Y. Sidis, J. Kulda, Z. Q. Mao, Y. Maeno, I. I. Mazin, and M. Braden, Spin fluctuations in Sr$_2$RuO$_4$ from polarized neutron scattering: implications for superconductivity, \href{https://doi.org/10.1103/PhysRevLett.122.047004}{Phys. Rev. Lett. {\bf 122}, 047004 (2019)}.

\bibitem{Kikugawa} N. Kikugawa, C. Bergemann, A. P. Mackenzie, and Y. Maeno, Band-selective modification of the magnetic fluctuations in ${\mathrm{Sr}}_{2}{\mathrm{RuO}}_{4}$: A study of substitution effects, \href{https://doi.org/10.1103/PhysRevB.70.134520}{Phys. Rev. B {\bf 70}, 134520 (2004)}.

\bibitem{Mravlje} J. Mravlje, M. Aichhorn, T. Miyake, K. Haule, G. Kotliar, and A. Georges, Coherence-Incoherence Crossover and the Mass-Renormalization Puzzles in ${\mathrm{Sr}}_{2}{\mathrm{RuO}}_{4}$, \href{https://doi.org/10.1103/PhysRevLett.106.096401}{Phys. Rev. Lett. {\bf 106}, 096401 (2011)}.

\bibitem{Kugler} F. B. Kugler, M. Zingl, H. U. R. Strand, S.-S. B. Lee, J. von Delft, and A. Georges, Strongly Correlated Materials from a Numerical Renormalization Group Perspective: How the Fermi-Liquid State of Sr$_2$RuO$_4$ Emerges,
\href{https://doi.org/10.1103/PhysRevLett.124.016401}{Phys. Rev. Lett. {\bf 124}, 016401 (2020)}.

\bibitem{CrTe2} K. Park, J.-E. Lee, D. Kim, Y. Zhong, C. Farhang, H. Lee, H. Im, W. Choi, S. Lee, S. Mun, et. al., Unusual Ferromagnetic Band Evolution and High Curie Temperature in Monolayer 1T-CrTe$_2$ on Bilayer Graphene, \href{https://doi.org/10.1002/smll.202506671}{Small, e06671 (2025)}. 

\bibitem{Wilhelm} A. Wilhelm, F. Lechermann, H. Hafermann, M. I. Katsnelson, and A. I. Lichtenstein, From Hubbard bands to spin-polaron excitations in the doped Mott material Na$_x$CoO$_2$, \href{https://doi.org/10.1103/PhysRevB.91.155114}{Phys. Rev. B {\bf 91}, 155114 (2015)}.

\bibitem{Katanin} A. A. Katanin, Splitting of electronic spectrum in paramagnetic phase of itinerant ferromagnets and altermagnets, \href{https://doi.org/10.48550/arXiv.2509.23396}{ArXiv:2509.23396}.





\bibitem{DTRILEX1} E. A. Stepanov, V. Harkov, and A. I. Lichtenstein, Consistent partial bosonization of the extended Hubbard model, \href{https://doi.org/10.1103/PhysRevB.100.205115}{Phys. Rev. B {\bf 100}, 205115 (2019)}.

\bibitem{DTRILEX2} V. Harkov, M. Vandelli, S. Brener, A. I. Lichtenstein, and E. A. Stepanov, Impact of partially bosonized collective fluctuations on electronic degrees of freedom, \href{https://doi.org/10.1103/PhysRevB.103.245123}{Phys. Rev. B {\bf 103}, 245123 (2021)}.

\bibitem{DTRILEX3} M. Vandelli, J. Kaufmann, M. El-Nabulsi, V. Harkov, A. I. Lichtenstein, and E. A. Stepanov, Multi-band D-TRILEX approach to materials with strong electronic correlations, \href{https://doi.org/10.21468/SciPostPhys.13.2.036}{SciPost Phys. {\bf 13}, 036 (2022)}. 
\bibitem{DGA} A. Toschi, A. A. Katanin, and K. Held, Dynamical vertex approximation: A step beyond dynamical mean-field theory, \href{https://doi.org/10.1103/PhysRevB.75.045118}{Phys. Rev. B {\bf 75}, 045118 (2007)}.

\bibitem{DGA1} A. A. Katanin, A. Toschi, K. Held, Comparing pertinent effects of antiferromagnetic fluctuations in the two and three dimensional Hubbard model, \href{https://doi.org/10.1103/PhysRevB.80.075104}{Phys. Rev. B {\bf 80}, 075104 (2009)}.

\bibitem{SCDGA} J. Kaufmann, C. Eckhardt, M. Pickem, M. Kitatani, A. Kauch, and K. Held, Self-consistent ladder D$\Gamma$A approach, \href{https://doi.org/10.1103/PhysRevB.103.035120}{Phys. Rev. B {\bf 103}, 035120 (2021)}.

\bibitem{PhysRevB.77.033101} A. N. Rubtsov, M. I. Katsnelson, and A. I. Lichtenstein, Dual fermion approach to nonlocal correlations in the Hubbard model, \href{https://doi.org/10.1103/PhysRevB.77.033101}{Phys. Rev. B {\bf 77}, 033101 (2008)}.

\bibitem{BRENER2020168310} S. Brener, E. A. Stepanov, A. N. Rubtsov, M. I. Katsnelson, and A. I. Lichtenstein, Dual fermion method as a prototype of generic reference-system approach for correlated fermions, \href{https://doi.org/10.1016/j.aop.2020.168310}{Ann. Phys. {\bf 422}, 168310 (2020)}.

\bibitem{Rubtsov20121320} A. N. Rubtsov, M. I. Katsnelson, and A. I. Lichtenstein, Dual boson approach to collective excitations in correlated fermionic systems, \href{https://doi.org/10.1016/j.aop.2012.01.002}{Ann. Phys. {\bf 327}, 1320 (2012)}.

\bibitem{PhysRevB.90.235135} E. G. C. P. van Loon, A. I. Lichtenstein, M. I. Katsnelson, O. Parcollet, and Hartmut Hafermann, Beyond extended dynamical mean-field theory: Dual boson approach to the two-dimensional extended Hubbard model, \href{https://doi.org/10.1103/PhysRevB.90.235135}{Phys. Rev. B {\bf 90}, 235135 (2014)}.

\bibitem{PhysRevB.93.045107} E. A. Stepanov,  E. G. C. P. van Loon, A. A. Katanin, A. I. Lichtenstein, M. I. Katsnelson, and A. N. Rubtsov, Self-consistent dual boson approach to single-particle and collective excitations in correlated systems, \href{https://doi.org/10.1103/PhysRevB.93.045107}{Phys. Rev. B {\bf 93}, 045107 (2016)}.

\bibitem{PhysRevB.100.165128} L. Peters, E. G. C. P. van Loon, A. N. Rubtsov, A. I. Lichtenstein, M. I. Katsnelson, and E. A. Stepanov, Dual boson approach with instantaneous interaction, \href{https://doi.org/10.1103/PhysRevB.100.165128}{Phys. Rev. B {\bf 100}, 165128 (2019)}.





\bibitem{PhysRevLett.127.207205}
E. A. Stepanov, Y. Nomura, A. I. Lichtenstein, and S. Biermann, Orbital Isotropy of Magnetic Fluctuations in Correlated Electron Materials Induced by Hund's Exchange Coupling, \href{https://doi.org/10.1103/PhysRevLett.127.207205}{Phys. Rev. Lett. {\bf 127}, 207205 (2021)}.

\bibitem{stepanov2021coexisting} E. A. Stepanov, V. Harkov, M. R{\"o}sner, A. I. Lichtenstein, M. I. Katsnelson, and A. N. Rudenko, Coexisting charge density wave and ferromagnetic instabilities in monolayer InSe, \href{https://doi.org/10.1038/s41524-022-00798-4}{npj Comput. Mater. {\bf 8}, 118 (2022)}.

\bibitem{PhysRevLett.129.096404} E. A. Stepanov, Eliminating Orbital Selectivity from the Metal-Insulator Transition by Strong Magnetic Fluctuations, \href{https://doi.org/10.1103/PhysRevLett.129.096404}{Phys. Rev. Lett. {\bf 129}, 096404 (2022)}.

\bibitem{PhysRevResearch.5.L022016} M. Vandelli, J. Kaufmann, V. Harkov, A. I. Lichtenstein, K. Held, and E. A. Stepanov, Extended regime of metastable metallic and insulating phases in a two-orbital electronic system, \href{https://doi.org/10.1103/PhysRevResearch.5.L022016}{Phys. Rev. Res. {\bf 5}, L022016 (2023)}.

\bibitem{PhysRevLett.132.226501} E. A. Stepanov and S. Biermann, Can Orbital-Selective N\'eel Transitions Survive Strong Nonlocal Electronic Correlations?, \href{https://doi.org/10.1103/PhysRevLett.132.226501}{Phys. Rev. Lett. {\bf 132}, 226501 (2024)}.

\bibitem{PhysRevB.110.L161106} E. A. Stepanov, M. Chatzieleftheriou, N. Wagner, and Giorgio Sangiovanni, Interconnected renormalization of Hubbard bands and Green's function zeros in Mott insulators induced by strong magnetic fluctuations, \href{https://doi.org/10.1103/PhysRevB.110.L161106}{Phys. Rev. B {\bf 110}, L161106 (2024)}.

\bibitem{stepanov2023charge} E. A. Stepanov, M. Vandelli, A. I. Lichtenstein, and F. Lechermann, Charge Density Wave Ordering in NdNiO$_2$: Effects of Multiorbital Nonlocal Correlations, \href{https://doi.org/10.1038/s41524-024-01298-3}{npj Comput. Mater. {\bf 10}, 108 (2024)}.

\bibitem{j6bj-gz7j} E. A. Stepanov, Fingerprints of a charge ice state in the doped Mott insulator Nb$_3$Cl$_8$, \href{https://doi.org/10.1103/j6bj-gz7j}{Phys. Rev. B {\bf 112}, 045131 (2025)}.

\bibitem{arXiv.2502.08635} E. A. Stepanov, S. Iskakov, M. I. Katsnelson, and A. I. Lichtenstein, Superconductivity of Bad Fermions: Origin of Two Gaps in HTSC Cuprates, \href{https://doi.org/10.48550/arXiv.2502.08635}{ArXiv:2502.08635}.

\bibitem{CT-QMC1} A. N. Rubtsov, V. V. Savkin, and A. I. Lichtenstein, Continuous-time quantum Monte Carlo method for fermions, \href{https://doi.org/10.1103/PhysRevB.72.035122}{Phys.
Rev. B \textbf{72}, 035122 (2005)}.

\bibitem{CT-QMC2} P. Werner, A. Comanac, L. de Medici, M.
Troyer, and A. J. Millis, Continuous-Time Solver for Quantum Impurity Models, \href{https://doi.org/10.1103/PhysRevLett.97.076405}{Phys. Rev. Lett. \textbf{97}, 076405 (2006)}.

\bibitem{iQIST} Li Huang, Y. Wang, Zi Yang Meng, L. Du, P. Werner, and Xi Dai, {iQIST: An open source continuous-time quantum Monte Carlo impurity solver toolkit}, \href{http://dx.doi.org/10.1016/j.cpc.2015.04.020}{Comp. Phys. Comm. \textbf{195}, 140 (2015)}; Li Huang, {iQIST v0.7: An open source continuous-time quantum Monte Carlo impurity solver toolkit},  \href{http://dx.doi.org/10.1016/j.cpc.2017.08.026}{Comp. Phys. Comm. \textbf{221}, 423 (2017)}.

\bibitem{ana_cont} J. Kaufmann and K. Held, ana\_cont: Python package for analytic continuation, \href{https://doi.org/10.1016/j.cpc.2022.108519}{Comput. Phys. Commun. {\bf 282}, 108519 (2023)}.

\bibitem{fRG2011} A. A. Katanin, H. Yamase, and V. Yu. Irkhin, Ferromagnetic instability and finite-temperature properties of two-dimensional electron systems with van Hove singularities, \href{https://doi.org/10.1143/JPSJ.80.063702}{J. Phys. Soc. Jpn. {\bf 80}, 063702 (2011)}.

\bibitem{My_BS} A. A. Katanin, Improved treatment of fermion-boson vertices and Bethe-Salpeter equations in non-local extensions of dynamical mean field theory, \href{https://doi.org/10.1103/PhysRevB.101.035110}{Phys. Rev. B \textbf{101}, 035110 (2020)}.

\bibitem{FirstFreq} A. V. Chubukov and D. L. Maslov, First-Matsubara-frequency rule in a Fermi liquid. I. Fermionic self-energy, \href{https://doi.org/10.1103/PhysRevB.86.155136}{Phys. Rev. B {\bf 86}, 155136 (2012)}.


\bibitem{Khodel} V. A. Khodel and V. R. Shaginyan, Superfluidity in system with fermion condensate, \href{http://jetpletters.ru/ps/1143/article_17312.shtml}{JETP Lett. {\bf 51}, 553 (1990)}; G. E. Volovik, A new class of normal Fermi liquids, \href{http://jetpletters.ru/ps/1149/article_17392.shtml}{JETP Lett. {\bf 53}, 222 (1991)}; M.
V. Zverev and M. Baldo, The multi-connected momentum distribution and fermion condensation, \href{https://doi.org/10.1088/0953-8984/11/9/004}{J. Phys.: Condens. Matter {\bf 11}, 2059 (1999)}.

\bibitem{Dzyaloshinskii} I. E. Dzyaloshinskii, Extended Van-Hove Singularity and Related Non-Fermi Liquids, \href{https://doi.org/10.1051/jp1:1996127}{J. Phys. I (France) {\bf 6}, 119 (1996)}.

\bibitem{RobvH} V. Yu. Irkhin, A. A. Katanin, and M. I. Katsnelson, Robustness of the van Hove scenario for high-Tc superconductors, \href{https://doi.org/10.1103/PhysRevLett.89.076401}{Phys. Rev. Lett. {\bf 89}, 076401 (2002)}.







\end{thebibliography}
\end{document}